\documentclass[10pt,journal,compsoc]{IEEEtran}
\ifCLASSOPTIONcompsoc
  \usepackage[nocompress]{cite}
\else
  \usepackage{cite}
\fi

\ifCLASSINFOpdf
  \usepackage{graphicx}
\else
\fi

\usepackage[cmex10]{amsmath}
\usepackage[linesnumbered]{algorithm2e}
\usepackage{color}
\ifCLASSOPTIONcompsoc
\usepackage[tight,normalsize,sf,SF]{subfigure}
\else
\usepackage[tight,footnotesize]{subfigure}
\fi

\usepackage{caption}
\usepackage{color}
\usepackage{ragged2e}
\usepackage{float}
\usepackage{array}

\usepackage{multirow}
\usepackage{url}
\ifCLASSINFOpdf
\else
\fi
\begin{document}
%
\title{ ACSEE: Antagonistic Crowd Simulation Model with Emotional Contagion
and Evolutionary Game Theory}
%
%
%
%


\author{
        Chaochao~Li,
        Pei~Lv,
        Dinesh~Manocha,
        Hua~Wang,
        Yafei~Li,
        Bing~Zhou,
        and Mingliang~Xu*


 \IEEEcompsocitemizethanks{
 \IEEEcompsocthanksitem *Corresponding author.
 \IEEEcompsocthanksitem Chaochao Li, Pei Lv, Hua Wang, Yafei Li, Bing Zhou, and Mingliang Xu are with Center for Interdisciplinary Information Science Research, ZhengZhou University, 450000.
 \IEEEcompsocthanksitem Dinesh Manocha is with Department of Computer Science and Electrical $\&$ Computer Engineering, University of Maryland, College Park, MD, USA.\protect\\
 E-mail: \{ielvpei, ieyfli, iebzhou, iexumingliang\} @zzu.edu.cn;  \hfil\break zzulcc@gs.zzu.edu.cn; dm@cs.unc.edu; wanghua@zzuli.edu.cn
}
\thanks{}

}

%
%

\markboth{Journal of \LaTeX\ Class Files,~Vol.~X, No.~X, November~2019}%
{Shell \MakeLowercase{\textit{et al.}}: Bare Demo of IEEEtran.cls for Computer Society Journals}
%



\IEEEtitleabstractindextext{%
\begin{abstract}
\justifying
Antagonistic crowd behaviors are often observed in cases of serious conflict.
Antagonistic emotions, which is the typical psychological state of agents in different roles (i.e. cops, activists, and civilians) in crowd violent scenes, and the way they spread through contagion in a crowd are important causes of crowd antagonistic behaviors.
Moreover, games, which refers to the interaction between opposing groups adopting different strategies to obtain higher benefits and less casualties, determine the level of crowd violence.
We present an antagonistic crowd simulation model, ACSEE, which is integrated with antagonistic emotional contagion and evolutionary game theories. Our approach models the antagonistic emotions between agents in different roles using two components: mental emotion and external emotion. We combine enhanced susceptible-infectious-susceptible (SIS) and game approaches to evaluate the role of antagonistic emotional contagion in crowd violence. Our evolutionary game theoretic approach incorporates antagonistic emotional contagion through deterrent force, which is modelled by a mixture of emotional forces and physical forces defeating the opponents. Antagonistic emotional contagion and evolutionary game theories influence each other to determine antagonistic crowd behaviors. We evaluate our approach on real-world scenarios consisting of different kinds of agents. We also compare the simulated crowd behaviors with real-world crowd videos and use our approach to predict the trends of crowd movements in violence incidents. We investigate the impact of various factors (number of agents, emotion, strategy, etc.) on the outcome of crowd violence. We present results from user studies suggesting that our model can simulate antagonistic crowd behaviors similar to those seen in real-world scenarios.
\end{abstract}

\begin{IEEEkeywords}
Group violence, emotional contagion, evolutionary game theory
\end{IEEEkeywords}}

\maketitle

\IEEEdisplaynontitleabstractindextext

%
\IEEEpeerreviewmaketitle

\IEEEraisesectionheading{\section{Introduction}\label{introduction}}



Crowd simulation has received increased attention in virtual reality, games, urban modeling, and pedestrian dynamics.
One of the most important tasks in crowd simulation is to generate realistic crowd behaviors.
Physical methods\cite{102,113,051}, psychology principles\cite{107,014,034}, or approaches from other relatively matured disciplines \cite{114,115,0521,079} are leveraged into the crowd simulation to improve the similarity between simulation results and real-world crowd movements.
As pointed out in \cite{107}, emotion has a great influence on crowd behavior and it often invokes an agent to implement either a positive or negative behavioral response. Thus, the emotion modeling in crowd simulation is always the main focus in latest research work.
However, the emotional aspect of antagonistic crowd behaviors among people in different roles is left unexplored \cite{108}.
Analyzing the emotions of antagonistic crowd behaviors is indeed extremely important, as it can help us understand evolution process of antagonistic crowd behaviors and predict trends of crowd movements.

In this paper, we mainly deal with the problem of simulating antagonistic crowd behaviors. Such behaviors are associated with acts of violation and destruction and are typically carried out as a sign of defiance
against a central authority or an indication of conflict between opposing groups \cite{027}.
Our goal is to develop a new crowd simulation model that can predict trends of crowd movement in these situations while ignoring the trajectory of a particular individual, discuss the conditions of winning and losing sides, and help to develop measures to quell incidents of crowd violence.
Not only will such a method be useful for training police officers, but it could also predict the trends of crowd movements and provide the decision for controlling crowd violence incidents.


It is difficult to simulate realistic antagonistic crowd behaviors because of complex influencing factors. In practice, such behaviors are closely related to antagonistic emotions \cite{094,2019111408}, i.e. the emotions between opposed groups, and evolutionary game theory \cite{014,027}.
In the pursuit of more realistic antagonistic behaviors in virtual agents, antagonistic emotion simulation should be incorporated into crowd simulation models \cite{034}.
Most prior crowd simulation models ignore antagonistic emotions and individuals' antagonistic behaviors.
Fu et al. \cite{009} focus on agents' emotions for only one role without involving the antagonistic emotions between different types of agents.
Some empirical methods of modeling antagonistic behaviors are presented in the form of riot games \cite{067}, game theoretic models \cite{068}, and social networks. These methods are based on statistical spatial-temporal analysis and role-playing dynamics in crowds and can generate emergent social phenomena. Other models use evolutionary game theory to simulate the behaviors and interactions between different kinds of agents \cite{027,035}.
Evolutionary game theory has successfully helped explain many complex and challenging aspects
of biological and social phenomena in recent decades \cite{035}.
Inspired by the idea that more offspring will be produced by more fit biological organisms in a given environment,
evolutionary game theory provides us with the methodology
to study strategic interactions among agents in  incidents of crowd violence \cite{036}.
There is considerable work on evaluating individuals' emotions \cite{005,072}.
Panic, for example, destroys an individual's normal mental function,
transforms the individual into an irrational state, and can lead to unpredictable abnormal behaviors. 
Furthermore, Durupinar et al. \cite{014} point out that agents' emotions dominate their decision-making process in games and other behaviors.
However, the relationship between antagonistic emotion and antagonistic crowd behavior has not been fully explored \cite{009}.
 It is challenging to accurately model antagonistic emotions among agents in different roles because the antagonism is complex and changes constantly and dynamically \cite{011}. Moreover, prior methods do not consider the effect of antagonistic emotion on evolutionary game theory despite the fact that emotion influences an agent's behavior significantly. For example, Lv et al. \cite{116} calculate emotions based on political viewpoints of individuals at political rallies. Their method doesn't involve evolutionary game theory or explore the relationship between evolutionary game theory and antagonistic emotion.

Recent advances in crowd simulation models attempt to simulate plausible human behavior by introducing psychological phenomena to virtual
agents \cite{034}.
Inspired by the psychological theory in \cite{012}, which shows
the characterization and simulation of emotional contagion, these psychological phenomena effectively improve the reliability of simulations.
Fu et al.\cite{009} integrate emotional contagion with individual movement to obtain realistic emotions and behaviors in a crowd during an emergency.
Based on this method, we propose an enhanced SIS model considering the benefits of games and test our antagonistic emotional contagion method among agents in different roles.
Moreover, we integrate antagonistic emotional contagion with evolutionary game theory through deterrent force, which describes the differences between agents more accurately. Finally, we determine antagonistic crowd behaviors combining antagonistic emotional contagion and evolutionary theories.

Our main contributions include:
\begin{itemize}
%

\item We propose a method to simulate emotion contagion between antagonistic agents in different roles, which is determined by combing the enhanced SIS and game approaches.
\item We propose a kind of deterrent force, which is modelled by a mixture of emotional forces and physical forces defeating the opponents, and it helps individuals make reasonable decisions in the games.
\item We integrate antagonistic emotions into an evolutionary game theoretic method through deterrent force to model antagonistic crowd behaviors more realistically.

\end{itemize}

\begin{figure*}[htb] \begin{centering}
  \centering
  \includegraphics[width=13.5cm]{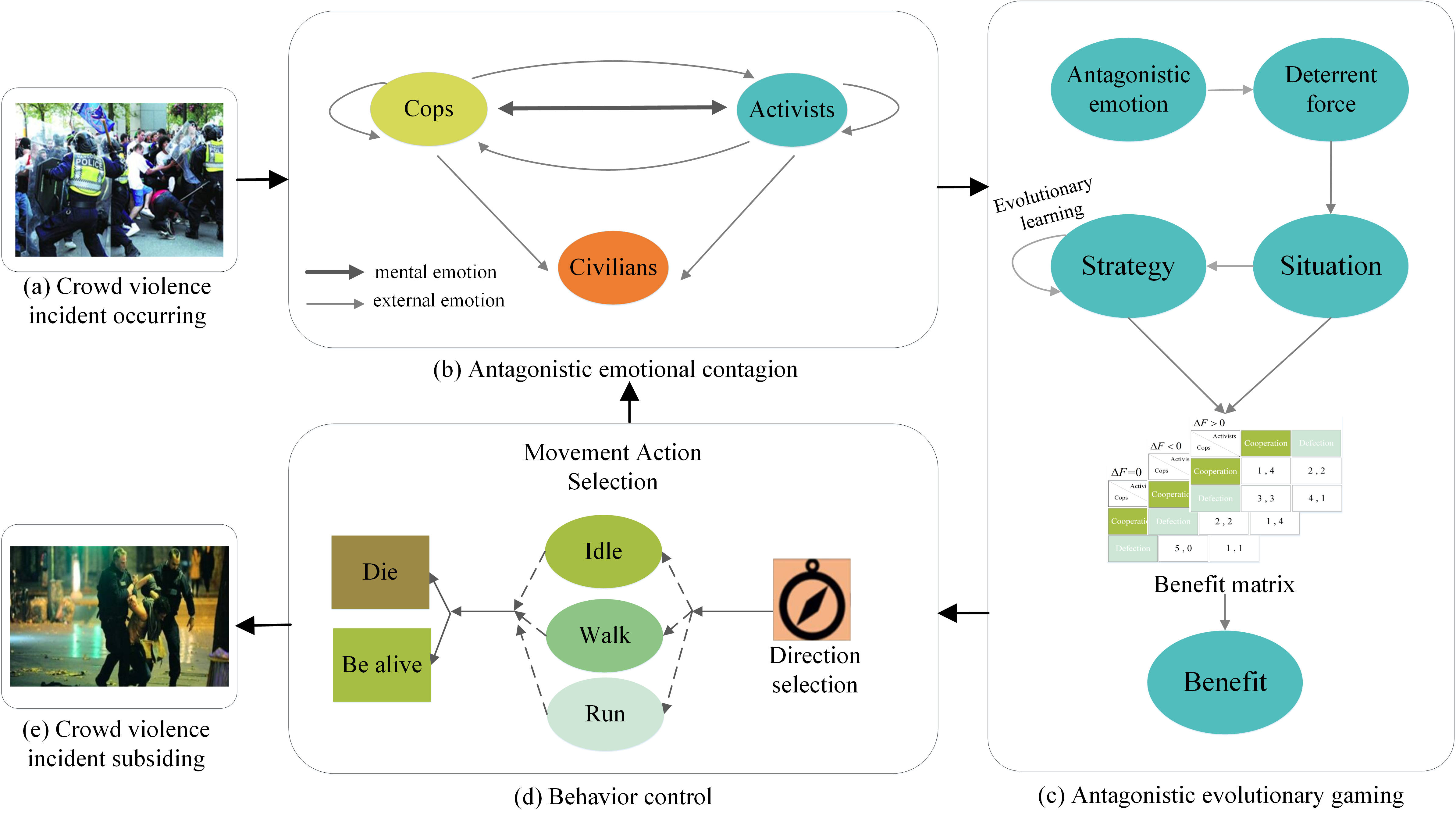}
  \centering
  \caption{Overview of our ACSEE model.
(a) Crowd violence occurs. Civilians, activists, and cops are specified in our model and they can perceive environmental information.
The basic attributes of these agents are introduced in Section \ref{Basic model based on multi-agent}.
(b) Our antagonistic emotional contagion method consists of two components: mental emotion and external emotion. After the outbreak of the crowd violence,
the antagonistic emotions of each agent are updated at each time step (Section \ref{Emotional contagion mechanism}).
(c) We calculate the antagonistic emotion and deterrent force of each agent.
The situation of the cops and activists is determined based on the deterrent force, which is defined according to antagonistic emotion.
The game between antagonistic cops and activists is carried out in different situations.
Agents use different strategies to carry out game interactions and their benefits are analyzed (Section \ref{Evolutionary game mechanism}).
Their strategies are evolved over time using the evolutionary learning method.
(d) We present behavior control method based on deterrent force, strategy, and benefit (Section \ref{Behavior mechanism}).
Agents choose rational movement directions and actions and their positions are updated.
We use step (b) to calculate the positions of all the agents at the next time step.
(e) If the crowd violence subsides, the incident ends.}
  \label{fig:1}
  \end{centering}
\end{figure*}

Our model can generate simulation results that are closest to the real-world scenarios in the overall trends of crowd movements.
We have implemented our model and tested it on several outdoor scenarios with varying ratios of different types of roles. Our simulation results are compared with the real-world videos and we evaluate the benefits of our model by performing user studies. The results indicate that the behaviors of agents generated by our model are closer to real-world scenes in the overall trend of crowd movements than those seen in other methods.

The rest of this paper is organized as follows. We review related work in Section \ref{related work}. We give an overview of our method in Section \ref{Preliminary}. We introduce our model in Section \ref{our model} and show it can be used to generate antagonistic crowd behaviors. We describe the implementation and highlight its performance on complex scenarios in Section \ref{experiments}. We also present results from our preliminary user studies in Section \ref{experiments}.

\section{Related Work}
\label{related work}

In this section, we provide a brief overview of prior works in emotional contagion, evolutionary game theory, and agent-based crowd simulation.

\subsection{Emotional contagion }

Emotion is a psychological parameter and has a significant influence on individuals in a crowd \cite{014,104,105,2019111201}.
Emotional contagion is closely related to human movement \cite{2019111405,2019111406,2019111407}.
In this subsection,
we introduce some representative works about emotional contagion \cite{107}.

The epidemiological susceptible-infectious-recovered (SIR) model \cite{015} divides the individuals in a crowd into three categories: infected, susceptible, and recovered.
The analysis of the spread of epidemic among these three groups has also been extended to other fields. In \cite{016},
the extended model is used to simulate the spread of rumors. Some researchers use the epidemiological SIR model in conjunction with
other models to describe emotion propagation under specific situations.
In \cite{014}, the epidemiological SIR model is improved by combining it with the OCEAN (openness, conscientiousness, extroversion, agreeableness, neuroticism) personality model \cite{054}. In \cite{019}, a qualitatively simulated approach to model emotional contagion is proposed for large-scale emergency evacuation. The method shows that the effectiveness of rescue guidance is influenced by the leading
emotions in a crowd. Moreover, in \cite{009}, the cellular automata model based on the SIR model (CA-SIRS) is used to describe emotional
contagion in a crowd during an emergency, capturing the dynamic process ``susceptible-infected-recovered-susceptible.''
However, in some case, people only need to consider two emotional states: infected and susceptible. Hill et al. \cite{2019102101} evaluate the spread of long-term emotional states across a social network based on the classical SIS (Susceptible Infected Susceptible) model.
Cai et al. \cite{20191021} combine the OCEAN and SIS models to simulate emotional contagion on crowd evacuation.
Song et al. \cite{2019102102} discuss the factors influencing individual evacuation decision making in the view of social contagion based on Susceptible-Infective (SI) model. Emotional contagion can also be used in traffic simulation \cite{2019111401,2019111402,2019111403} and crowd queuing simulation \cite{2019111404}.

A thermodynamic-based emotional contagion model is introduced by Bosse
et al. \cite{047} in the ASCRIBE system. The authors use a multi-agent based approach to define emotional contagion within groups.
Their study focuses on the emotions of a collective entity rather than the emotions of single individuals.
Neto et al. \cite{034} adapt the model proposed by Bosse et al. \cite{048} into BioCrowds and cope with different groups of agents.
Tsai et al. \cite{046} present an emotional contagion model that spreads the highest level of emotion to surrounding agents
in their ESCAPES framework.
In \cite{017}, dynamic emotion propagation is described from the
perspective of social psychology, combining thermodynamic-based models and epidemiological-based models.


According to the behaviors of different agents in antagonistic scenarios, we define three kinds of agents. Each kind of agents is assigned a role separately: civilians, activists, and cops.
Because of the antagonism between agents with different roles, the above-mentioned emotional contagion models can't be directly applied to crowd violent scenes.
In this paper, we propose antagonistic emotions. Antagonistic emotions are the opposite emotions of agents in different roles. Different roles of agents come from opposing groups in crowd violent scenes. In our model the emotions of cops and activists are the antagonistic emotions.
Antagonistic emotional contagion is the emotional contagion process of antagonistic agents in different roles under certain situations.
Our antagonistic emotional contagion method can quantitatively characterize dynamic changes in the antagonistic emotions between different roles.

\subsection{Evolutionary game theory}
In this subsection, we summarize some representative works about evolutionary game theory.

Evolutionary game theory has solved many biological problems.
For example, in \cite{041}, genome-driven evolutionary game theory helps to explain the rise of metabolic interdependencies in microbial communities.

Some researchers have applied evolutionary game theory to the recommendation system.
Saab et al. \cite{037} use classical and spatial evolutionary game theory as a possible solution to the Sybil
attack in recommender systems.
Li et al. \cite{042} propose a new stability analysis of repeated games and evolutionary games based on a subset of Nash equilibrium.

Evolutionary game theory can help us understand the behaviors of individuals better.
Huang et al. \cite{070} present a model in which every mutation leads to a new game between the mutants and the residents based on a evolutionary game theoretic approach.
Evolutionary game theory is one of the most effective approaches to understand and analyze widespread
cooperative behaviors among individuals \cite{039}.
 In \cite{040}, the combination of evolutionary game theory and graph theory provides an extended
framework to investigate cooperative behavior in social systems.
Quek et al. \cite{027} focus on the development of a spatial evolutionary
multiagent social network to study the macroscopic-behavioral dynamics of civil violence due to microscopic
game-theoretic interactions between goal-oriented agents.


Some previous works \cite{027,083} study antagonistic crowd behaviors in crowd violence incidents based on evolutionary game theory
but do not fully consider the differences of antagonistic emotions and deterrent forces of agents with different roles.
In this paper, we further improve the evolutionary game theoretic method.
At first, game in our model is established between opposing groups (cops and activists). When these agents confront different scenarios and situations, they adopt different strategies such as defection or cooperation to get higher benefits and less casualties.
Evolutionary game theory is a modeling approach of strategic interactions among agents in crowd violent incidents based on natural selection mechanism. Natural selection mechanism means that more offspring will be produced by more fit biological organisms in a given environment. We use evolutionary game theory to analyze the strategies and benefits of agents.
Our model incorporates antagonistic emotions into evolutionary game theory to describe the differences between agents, which estimates situations of antagonistic scenes more accurately.

%
%

\subsection{Agent-based crowd simulation}
Agent-based model is a kind of versatile method can simulate complex scenarios. In an agent-based model, all the agents are endowed with greater autonomy and have their own inherent attributes and properties. Agents can receive information from the surrounding environment, which will influence their actions and decisions \cite{81637}. They have separate velocities and moving directions. Hence, an agent-based model can produce complex crowd behaviors. In contrast to agent-based models, flow-based and particle-based models cannot accurately describe the differences between agents \cite{81638}. A flow-based model is mainly used to simulate high-density crowds and has no individuals or groups. A particle-based model cannot model high-level decision-making behaviors. Therefore, we integrate the proposed model with an agent-based method. In this subsection, we summarize this kind of methods.


An agent-based approach is the most common way to simulate crowd movements \cite{8169}. Kountouriotis et al. \cite{81638} use the agent-based model to simulate thousands of agents in real time, which integrates a high level of individual parametrization, such as group behaviors between friends and between a leader and a follower. Luo et al. \cite{8165} introduce a novel framework for proactive steering in agent-based crowd simulation. The Social Forces Model is combined with an agent-based method to simulate crowds \cite{81611}. In this model, repulsive and tangential forces of each agent are introduced to avoid collisions with surrounding agents and obstacles. However, all the agents share the same attributes and move with the same speed, which doesn't conform to real-world scenarios.

Some agent-based methods are used to simulate crowd movements in emergency scenarios. In these scenarios, the emotional state of an agent is a very important influencing factor in simulating realistic behaviors.
Shendarkar et al. \cite{8161} present a novel crowd simulation model for emergency response using BDI (belief, desire, intention) agents.
Luo et al. \cite{8162} describe the human-like decision-making process based on various physiological, emotional, and social group attributes for agents under normal and emergency situations. Aydt et al. \cite{8163} propose an emotion model integrated with an agent-based method in serious games based on modern appraisal theory. In contrast to the above methods, which don't involve antagonistic scenes or emotions between agents in different roles, our method focuses on antagonistic emotions and the relationship between antagonistic emotions and evolutionary game theory.


\begin{table}[htb]\scriptsize
\setlength{\belowcaptionskip}{10pt}
\renewcommand{\arraystretch}{1.3}
\caption{The parameters used in our ACSEE model.}
\label{terminology and parameters}
\centering
\begin{tabular}{m{2.3cm}<{\centering}|m{5.5cm}}
\hline
\bfseries Notation & \bfseries Description\\
\hline\hline
 \textit{PR} &  The radius of perceived range \\
\hline
 $e_{i}^{ex}$ & External emotion of agent $i$  \\
\hline
 $e_{i}^{me}$  & Mental emotion of agent $i$ \\
\hline
  $\Delta {{e}_{i,j}^{ex}}\left( t \right)$ & The increase in the strength of agent $i$'s external emotion received from agent $j$ at time $t$  \\
\hline
 $\Delta e_{c}^{ex}\left( t \right)$  & The increment of external emotion of agent $c$  at time $t$  \\
\hline
 $\Delta ben{{e}_{i}}\left( t \right)$  & The difference of the benefits of the games at time $t$ and $t-1$ for agent $i$\\
\hline
 $\Delta e_{i}^{me}\left( t \right)$ & The increment of the mental emotion of agent $i$ at time $t$ \\
\hline
 $\Delta {{e}_{i}}\left( t \right)$ & The increment of the total emotion of agent $i$ at time $t$  \\
\hline
  ${{e}_{i}}\left( t \right)$ & The total emotion of agent $i$ at time $t$  \\
\hline
 $T_{a2c}$ & If the emotion value of an activist exceeds the threshold $T_{a2c}$, role transition from activist to civilian occurs. \\
\hline
 $T_{c2a}$ & If the emotion value of a civilian less than the threshold $T_{c2a}$ , role transition from civilian to activist occurs. \\
\hline
 ${{f}_{i}}\left( t \right)$ & The deterrent force of agent $i$ at time $t$\\
\hline
 ${{F}_{i}}\left( t \right)$ & The total deterrent force of agents of the same type for agent $i$ at time $t$ \\
\hline
 ${{\hat{F}}_{i}}\left( t \right)$ & The total deterrent force of his or her opponents in the cells neighboring agent $i$ at time $t$ \\
\hline
 $\Delta {{F}_{i}}$ & The difference of the total deterrent forces between cops and activists that agent $i$ can perceive at time $t$ \\
\hline
 ${{P}_{die}}$ & The death probability  \\
\hline
  $T_{warn}$ & The early warning threshold \\
\hline
  $warn\_time$ & The time of early warning \\
\hline
 $T_{warn\_time}$ & The time of early warning threshold \\
\hline
\end{tabular}
\end{table}

\section{Overview of our approach} \label{Preliminary}
In this section, we introduce some basic and important concepts about crowd behavior simulation and crowd emotional contagion. We also give an overview of our method.

\subsection{Crowd behavior simulation}

Crowd behavior simulation can be defined as a process of emulating or simulating the movement of large amount of entities, characters or agents \cite{111}. At a broad level, crowd movement is governed by psychological status of individuals and their surrounding environment \cite{011}.
When humans form a crowd, interaction becomes an essential part of the overall crowd movement \cite{014}.
For agent-based methods of crowd simulation used in this paper, each agent is assumed as an independent decision-making entity, which has knowledge of the environment and a desired goal position at each step of the simulation. The interactions between an agent with others or with the environment are often performed at a local level \cite{110}. A typical crowd simulation model can be defined as in Equation \ref{eq111}. $P^t$ represents the positions of all the agents in the scene at time $t$ and $P^{t+1}$ is the positions of all the agents at time $t+1$, which can be induced by crowd simulation model $f$.

\begin{equation}\label{eq111}
P^{t+1}=f(P^t)
\end{equation} 

\subsection{Crowd emotional contagion}

Crowd simulation research has recently taken a new direction for modeling emotion of individuals to generate believable, heterogeneous crowd behaviors. Emotion of an agent can greatly affect its ability to perceive, learn, behave, and communicate within the surrounding environment\cite{014}. The emotion owned by one agent provides information about other agents' behavioral intentions and modulates his or her behavioral decision-making process. Based on their appraisal of the environment, emotions of the agents in a crowd are updated dynamically at different time. In antagonistic scenes, such emotional changes become more obvious and play a vital role in crowd interaction behaviors. As shown in Equation \ref{eq1112}, the emotion values of all the agents $E^{t+1}$ at time $t+1$ can be computed according to their relative positions $P^t$ and emotions of the agents $E^t$ at time $t$.

\begin{equation}\label{eq1112}
E^{t+1}=g(E^t,P^t)
\end{equation} 

\subsection{Overview of our ACSEE model}

This paper mainly discusses the influence of antagonistic emotions on agents' behaviors, whose purpose is to compute and update the status of all the agents at different time steps according to their emotions and roles. The crowd emotion is fully integrated into crowd behavior simulation, also with considering the confrontation between agents with different roles, such as civilians, activists, and cops.
Given the positions $P^t$, the emotions $E^t$, and the roles $R^t$ of all the agents at time $t$, our crowd simulation model $\text{ACSEE}(P^t,E^t)$ estimates the positions $P^{t+1}$, the emotions $E^{t+1}$, and the roles $R^{t+1}$ of all the agents at next time step as Equation~\ref{eq1113}.

\begin{equation}\label{eq1113}
\begin{aligned}
\{P^{t+1},E^{t+1},R^{t+1}\}&=\text{ACSEE}(f(P^{t-1}),g(E^{t-1},P^{t-1}),R^t)\\&=\text{ACSEE}(P^t,E^t,R^t)
\end{aligned}
\end{equation} 

\section{ACSEE Model}
\label{our model}

We present a novel \textbf{\emph{A}}ntagonistic \emph{\textbf{C}}rowd behavior \emph{\textbf{S}}imulation model (ACSEE) based on \emph{\textbf{E}}motional contagion and \emph{\textbf{E}}volutionary game theories. Our model consists of three important modules:
antagonistic emotional contagion, antagonistic evolutionary gaming, and behavior control. The antagonistic emotional contagion method
is designed by combining the enhanced SIS and game approaches.
Using the antagonistic emotions of agents,
we define their deterrent forces in Section \ref{Evolutionary game mechanism}. The enhanced evolutionary game theoretic approach is determined based on the deterrent forces of agents.
Our ACSEE model computes the behavior of each agent by modeling the influence from antagonistic emotional contagion and evolutionary game theories.
The flowchart of our ACSEE model is presented in Figure \ref{fig:1}.

\subsection{Symbols and Notations} \label{Symbols and Notation}

For convenience, the important parameters used in the ACSEE model  and their descriptions  are listed in Table \ref{terminology and parameters}.

\subsection{Agent modeling in antagonistic scenes} \label{Basic model based on multi-agent}

In this section, we mainly describe the role of different kinds of agents and the assumptions we formulated.

\subsubsection{Civilians, activists, and cops} \label{Civilians, activists, and cops}

Crowd violence incidents are often caused by some serious social contradictions, where a certain amount of activists challenge or break the normal and peaceful social order or stability in different ways of violence such as large-scale gathering, group activities, and physical conflicts.  We classify the agents in the crowd  under such situations as civilians, activists, or cops
based on their roles according to \cite{027,086}.

Civilians are neutral agents in the environment and pose no danger to the central authority. In general, civilians are vulnerable groups and do not participate in confrontation. The cops do their best to protect civilians while the activists persecute them.
 However, civilians may change their roles if conditions are favorable
 to express their anger and frustration publicly. For example, because of the instigation of the surrounding activists, the civilians may turn into activists to participate in the riot. Activists aim to create havoc and fuel the ongoing unrest while avoiding being defeated by cops. Cops maintain public order by suppressing activists and play a key role in preventing terrorist attack. In real-world scenarios, \emph{cops} and \emph{activists} can represent any two antagonistic groups \cite{094}. Civilians can also represent onlookers and neutral parties.


\subsubsection{Assumptions} \label{Assumptions}

Antagonistic crowds arise for many complex influencing factors. The simulation of antagonistic crowd behaviors considering all the influencing factors is an insoluble problem. From the observations of the real antagonistic crowd behaviors, we formulate the following assumptions to make this problem solvable.

\begin{itemize}
\item Emotions of cops are positive while those of activists are negative. Civilians are neutral agents. The process of emotional contagion can change their emotions. Cops with high positive emotions will make the agents around them more positive. Activists with high negative emotions will make the agents around them more negative. Civilians don't actively affect the emotions of surrounding agents \cite{014}.
\item An agent can maintain a perceived range that is centered around it.
We regard the perceived range of an agent as a circular area with a fixed radius $\textit{PR}$ \cite{088,014}.
\item Agents use different strategies to interact with their opposing agents. Agents' strategies refer to the actions taken during the games. According to the encountered situation, activists and cops can adopt one of two different strategies: cooperation or defection \cite{027,089}.
The cooperation strategy of activists means accepting peaceful settlements and running away from the cops.
The activists with defection strategy will revolt aggressively and instigate civilians to revolt.
The cops with cooperation strategy keep away from large gatherings of activists to protect civilians.
The defection strategy of cops means pursuing activists.
\item The agent may be classified as death. Dead agents are subdued by their opponents and pose no threat any more. It doesn't mean biological death.
\end{itemize}

\subsection{Antagonistic emotional contagion module} \label{Emotional contagion mechanism}

Since emotion has an important influence on people's behavior decision-making, accurate emotion modeling is essential and fundamental for a crowd simulation model\cite{014}.
In this section, we present our antagonistic emotional contagion module.
Emotions in our model incorporate the antagonism between agents in different roles.
${{e}_{i}}$ denotes the emotion of agent $i$.
${{e}_{i}}\in \left( -1, 1 \right)$ and ${e}_{i} \ne 0$.
There are two different types of emotion: positive emotion and negative emotion. When the emotion value is greater than 0,
the emotion is positive. The higher the emotion value of an agent, the more positive his/her emotion.
When the emotion value is less than 0, the emotion of the agent becomes negative. The lower the emotion value of an agent, the more negative his/her emotion. When an emotion value is closer to 0, the agent is regarded as being in a peaceful state and he or she tends to be conservative.
The descriptions of different emotion values of different roles are listed in Table \ref{The description of different emotion value of different roles}.

\begin{table}[htbp]\tiny
\setlength{\belowcaptionskip}{10pt}
\centering
\caption{The descriptions of different emotion values of different roles}
\label{The description of different emotion value of different roles}
\begin{tabular}{|c|c|c|c|c|}
\hline
\multicolumn{2}{|l|}{\multirow{2}{*}{Descriptions of emotion}} & \multicolumn{3}{c|}{Emotion}                                                                                                                                                                                         \\ \cline{3-5}
\multicolumn{2}{|l|}{}                                        & Closer to $1$                                                                            & Closer to $0$ & Closer to $-1$                                                                                                  \\ \hline
\multirow{3}{*}{Roles}               & Cops                   & \begin{tabular}[c]{@{}c@{}}high morale;\\ brave in subduing\\ the activists\end{tabular} & peaceful    & \begin{tabular}[c]{@{}c@{}}low morale;\\ afraid of activists;\\ dare not to fight with activists\end{tabular} \\ \cline{2-5}
                                     & Activists              & \begin{tabular}[c]{@{}c@{}}escape from cops \\ not challenging cops\end{tabular}       & peaceful    & arrogant and attack cops                                                                             \\ \cline{2-5}
                                     & Civilians              & \begin{tabular}[c]{@{}c@{}}brave;\\ not fear of activists\end{tabular}                 & peaceful    & \begin{tabular}[c]{@{}c@{}}uneasy, suffering, \\ fear of activists, and obeyed\end{tabular}                   \\ \hline
\end{tabular}
\end{table}

In such an antagonistic game scenario, individuals will be influenced by external and internal stimuli. External stimuli mainly come from the external environment and are often accepted by individuals passively. Internal stimuli come from the subjective perception and judgment of individuals by themselves. Both the internal and external stimuli in the antagonistic scenarios are able to produce emotions, so the emotion of an agent  ${{e}_{i}}$ consists of two parts.
The first part is the external
emotion $e_{i}^{ex}$ , which is influenced by surrounding agents. The second part is the mental emotion $e_{i}^{me}$ \cite{085}, which is determined by
an agent's own subjective consciousness. Therefore, the final emotion value is defined as follows \cite{118,119}:

\begin{equation}\label{eq1}
{{e}_{i}}=e_{i}^{ex}+e_{i}^{me}
\end{equation} 

\subsubsection{External emotion} \label{External emotion}

Our method for calculating external emotion is inspired by the emotional contagion model in \cite{009}.
An agent can be affected by others in his or her perceived range. The increase in the strength of
external emotion of agent $i$ ($\Delta {{e}_{i,j}^{ex}}\left( t \right)$) received from agent $j$ at time $t$ is defined as:

\begin{equation}\label{eq2}
\Delta {{e}_{i,j}^{ex}}(t)=[1-\frac{1}{(1+\exp (-L))}]\times {{e}_{i}}(t)\times {{A}_{j,i}}\times {{B}_{i,j}}
\end{equation} 
where $L$ represents the distance between agent $i$ and $j$, ${e}_{i}$ denotes the emotion of agent $i$,
${A}_{j,i}$ is the intensity of emotion received by $i$ from sender $j$,
and ${B}_{i,j}$ is the intensity of emotion which is sent from $i$ to receiver $j$.

Civilians can only passively receive the emotional contagion from surrounding agents and cannot actively influence others.
The increment of external emotion of agent $o$ at time $t$ is denoted as $\Delta e_{o}^{ex}\left( t \right)$.
$\Delta e_{o}^{ex}\left( t \right)$
includes emotional influences received from all the cops and activists in the perceived range of agent $o$ .
$\Delta e_{o}^{ex}\left( t \right)$
is defined as follows:

\begin{equation}\label{eq5}
\Delta e_{o}^{ex}(t)=\sum\limits_{i=1}^{k}{\Delta {{e}_{o,{{c}_{i}}}^{ex}}(t)+\sum\limits_{j=1}^{n}{\Delta {{e}_{o,{{a}_{j}}}^{ex}}}(t)}
\end{equation} 
where $\Delta {{e}_{o,{{c}_{i}}}^{ex}}\left( t \right)$ and $\Delta {{e}_{o,{{a}_{j}}}^{ex}}\left( t \right)$
denote the increase in the strength of the external emotion transmitted from cop ${c}_{i}$ and activist ${a}_{j}$ to agent $o$.

\subsubsection{Mental emotion} \label{Mental emotion}

Each agent establishes game play with agents in the opposing group
who are in his or her perceived range. Because civilians are neutral members and remain peaceful, we assume that no game interaction will occur between civilian agents. The benefits of each game are determined according to the method outlined in Section \ref{Evolutionary game mechanism}. Mental emotion is defined as the difference between the benefits
of two games and the mental emotion of civilians is a constant \cite{013}.

The difference between the benefits of the games at time $t$ and $t-1$ for agent $i$
is denoted as $\Delta ben{{e}_{i}}\left( t \right)=ben{{e}_{i}}\left( t \right)-ben{{e}_{i}}\left( t-1 \right)$.
The threshold that leads to emotional fluctuations is $\delta$. The relationship
between the increment of mental emotion and the difference between the benefits at time $t$ is defined by:

\begin{equation}\label{eq6}
\Delta e_{i}^{me}(t)=\left\{ \begin{matrix}
   \begin{matrix}
   rand(-0.01,0.01), & \left| \Delta {ben}{{{e}}_{i}}(t) \right|<\delta   \\
\end{matrix}  \\
   \begin{matrix}
   \frac{0.1}{\delta +\exp (\delta /\Delta ben{{e}_{i}}(t))}, & \Delta ben{{e}_{i}}(t)\ge \delta   \\
\end{matrix}  \\
   \begin{matrix}
   -\frac{0.1}{\delta +\exp (\Delta ben{{e}_{i}}(t)/\delta )}, & \Delta ben{{e}_{i}}(t)\le -\delta   \\
\end{matrix}  \\
\end{matrix} \right.
\end{equation} 

In Equation \ref{eq6}, $\left| \Delta ben{{e}_{i}}\left( t \right) \right|<\delta $ means that the difference between benefits
fails to reach the emotional fluctuation threshold $\delta $. Therefore, there is little change in the
emotion of agent $i$. In this case, the mental emotion value is a random number on the interval (-0.01, 0.01).
$\Delta ben{{e}_{i}}\left( t \right)\ge \delta $ means that the benefit at time $t$ is higher
than that at time $t-1$. The benefit increase makes the cops more positive and activists more negative.
$\Delta ben{{e}_{i}}\left( t \right)\le -\delta $ means that the difference between the
benefits of time $t$ and $t-1$ is higher than the emotional fluctuation threshold $\delta $. The benefit
at time $t$ is lower than that at time $t-1$. The benefit decrease makes cops more negative and activists more positive.

\subsubsection{Emotion updating} \label{Emotion updating}

\begin{figure}[htb] \begin{centering}
  \centering
  \includegraphics[width=9cm]{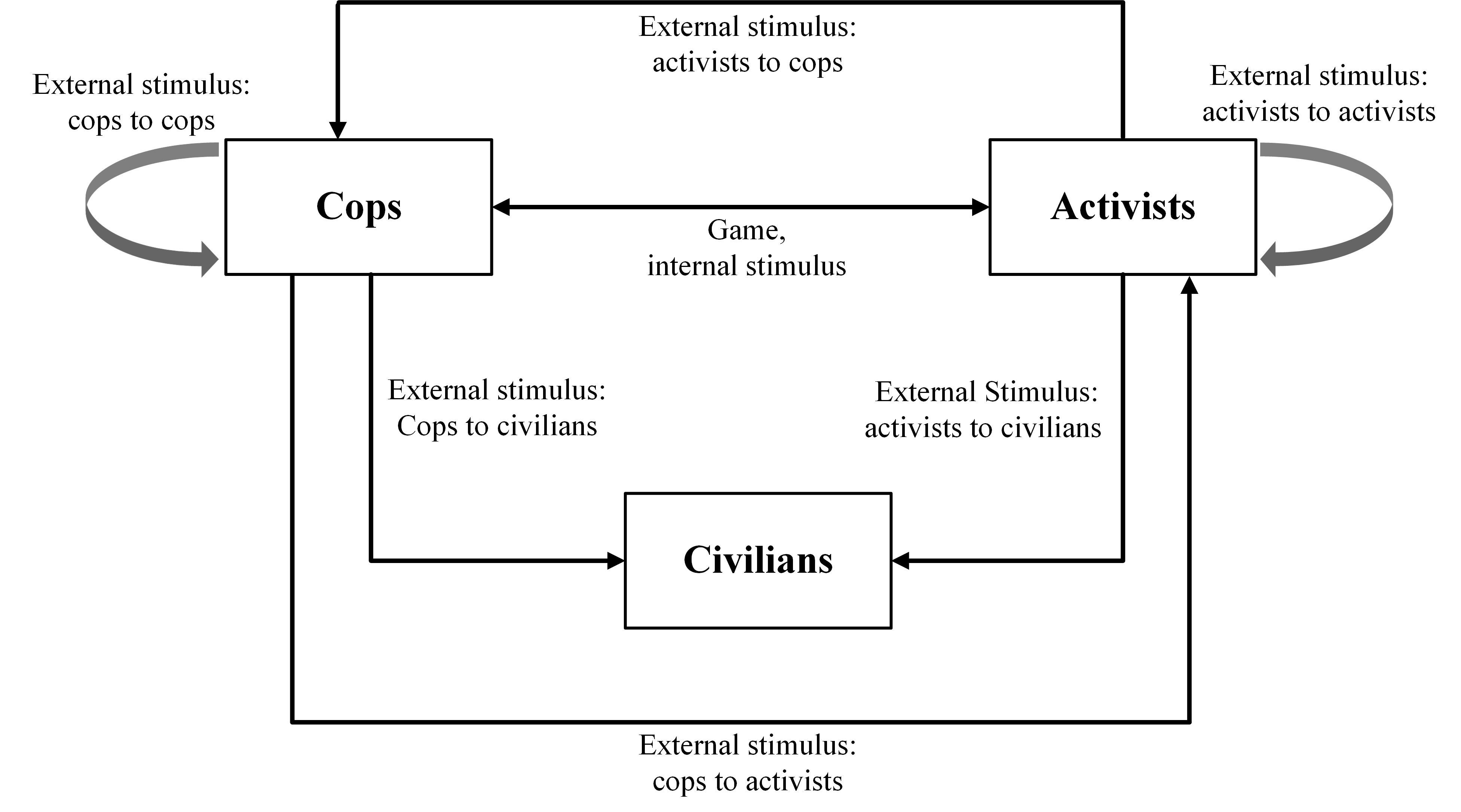}
  \centering
  \caption{ Emotion updating method. The benefit of the game between cops and activists is the internal stimulus, which leads to the changes in mental emotions. Emotional contagion among different kinds of agents is the external stimulus, which leads to the changes in external emotions.}
  \label{fig:2}
  \end{centering}
\end{figure}

Our emotion updating method for agents is presented in Figure \ref{fig:2}. The mental and external emotions of each agent are updated according to
the evolution of the games and changes in agents' locations. The external emotion of an agent is determined by the emotional contagion (external
stimulus) of surrounding cops and activists. The differences between the benefits of games (internal stimulus), which is defined in Section \ref{Mental emotion}, lead to the changes in the mental emotions of cops and activists.

The increment of the total emotion ($\Delta {{e}_{o}}\left( t \right)$) of agent $o$ at time $t$
 is defined as follows:

\begin{equation}\label{eq7}
\Delta {{e}_{o}}(t)=\Delta e_{o}^{ex}(t)+\Delta e_{o}^{me}(t)
\end{equation} 

For each time step, the total emotion value is updated. At time $t$, the total emotion value of agent $o$ is defined as follows:

\begin{equation}\label{eq10}
{{e}_{o}}(t)={{e}_{o}}(t-1)+\Delta {{e}_{o}}(t)
\end{equation} 

\subsection{Antagonistic evolutionary gaming module} \label{Evolutionary game mechanism}

In our model an agent establishes game play with all the agents from the opposing group within his or her perceived range, according to game theory. When agents confront different scenarios and situations, they adopt different strategies and get varying benefits. They aim to maximize their benefits and minimize casualties according to the current situation.
In this subsection, we present the evolutionary game theoretic module of our model, which is used to analyze the strategies and benefits of agents.

At first, agents estimate surrounding situations based
on the deterrent forces of the agents in their perceived range. Deterrent force is a kind of power by which an agent can beat his or her opponents and is closely related to the agent's behavior. Emotion plays a crucial role in agents' deterrent forces \cite{093}.
The deterrent force of agent $i$ is defined in the following equation:

\begin{equation}\label{eq13}
{{f}_{i}}\left( t \right)=\sin \left( \left| {{e}_{i}}\left( t \right) \right|\cdot \frac{\pi }{2} \right)
\end{equation} 
where ${{e}_{i}}\left( t \right)$ is the emotion of agent $i$ at time $t$. The more positive or negative the emotion of an agent is,
the greater the deterrent force the agent possesses \cite{093}. The total deterrent forces are defined as follows:

\begin{equation}\label{eq14}
{{F}_{i}}\left( t \right)=\underset{k\in A}{\mathop \sum }\,{{f}_{k}}\left( t \right)
\end{equation} 

\begin{equation}\label{eq15}
{{\hat{F}}_{i}}\left( t \right)=\underset{\hat{k}\in \hat{A}}{\mathop \sum }\,{{f}_{{\hat{k}}}}\left( t \right)
\end{equation} 
where the set $A$ denotes agents of the same type in the perceived range of agent $i$ and the set of the opposing agents in the perceived range of agent $i$ is $\hat{A}$.

The situation is defined according to the difference between the total deterrent forces of cops and activists perceived by agent $i$, which is expressed in Equation \ref{eq16}, as follows.

\begin{equation}\label{eq16}
\Delta {{F}_{i}}={{F}_{i}}\left( \text{t} \right)-{{\hat{F}}_{i}}\left( \text{t} \right)
\end{equation} 

Under different situations, the benefits gained during the games are different.
The benefit matrix is defined according to varying situations. In contrast, the benefit matrix
defined in \cite{027} is based on the number of cops and activists and assumes that
the deterrent forces of all the agents are the same. Instead, we define the {\em benefit matrix}
based on the deterrent forces of cops and activists. We fully account for the differences between the deterrent forces of all the agents,
which conforms to real-world scenes.
 The benefit matrix corresponding to different situations is shown in Table \ref{Payoff matrix of cops and activists in group violence}.

\renewcommand\arraystretch{1.5}
\begin{table}[htbp]\scriptsize
\setlength{\belowcaptionskip}{10pt}
\centering
\caption{Benefit matrix of cops and activists. $\Delta F$ denotes the situations of cops and activists.
$\Delta F>0$ means that the total deterrent force of the cops is higher than that of the activists.
$\Delta F<0$ and $\Delta F=0$ are defined in a similar way.
Activists and cops can adopt two different strategies: cooperation or defection.
We list the benefits of cops and activists corresponding to different situations. For example, the ratio of the benefits of the cops using a cooperation strategy to the activists using a cooperation strategy is 1:4 when $\Delta F>0$. The ratio of the benefits of the cops using a cooperation strategy to the activists using a defection strategy is 2:2 when $\Delta F>0$ \cite{027}}
\label{Payoff matrix of cops and activists in group violence}
\begin{tabular}{|c|c|c|c|c|}
\hline
\multicolumn{1}{|l|}{\multirow{2}{*}{Situations}} & \multicolumn{2}{c|}{\multirow{2}{*}{Benefit}}                                               & \multicolumn{2}{c|}{Strategy of activists} \\ \cline{4-5}
\multicolumn{1}{|l|}{}                                               & \multicolumn{2}{c|}{}                                                                       & Cooperation           & Defection          \\ \hline
\multirow{2}{*}{$\Delta F>0$}                                      & \multirow{6}{*}{\begin{tabular}[c]{@{}c@{}}Strategy\\ of\\ cops\end{tabular}} & Cooperation & 1,4                   & 2,2                \\ \cline{3-5}
                                                                     &                                                                               & Defection   & 3,3                   & 4,1                \\ \cline{1-1} \cline{3-5}
\multirow{2}{*}{$\Delta F<0$}                                         &                                                                               & Cooperation & 4,1                   & 3,3                \\ \cline{3-5}
                                                                     &                                                                               & Defection   & 2,2                   & 1,4                \\ \cline{1-1} \cline{3-5}
\multirow{2}{*}{$\Delta F=0$}                                                 &                                                                               & Cooperation & 3,3                   & 0,5                \\ \cline{3-5}
                                                                     &                                                                               & Defection   & 5,0                   & 1,1                \\ \hline
\end{tabular}
\end{table}

When the total deterrent force of the cops is higher than that of the activists ($\Delta \text{F}>0$),
if both groups adopt a strategy of cooperation, cops miss an opportunity to make arrests. Compared with activists, cops reap fewer benefits. If both groups defect, cops gain the upper hand because they have a higher total deterrent force and therefore reap more benefits. If cops cooperate and activists defect, both groups' benefits remain relatively neutral. Cops should defect to
confront activists while activists should cooperate to avoid challenging cops and inviting casualties. If cops defect and activists cooperate,
cops exert dominance over activists and activists avoid direct conflict. Therefore, both groups obtain benefits.
When the total deterrent force of cops is less than or equal to that of activists ($\Delta \text{F}<0$ or $\Delta \text{F}=0$),
their benefits are defined similarly.

After a game, the strategies of cops and activists are updated according to the results of that game.
Each agent is defined by a binary string. This string encodes the strategy bits in different situations.
This string suggests the strategies an agent should adopt when $\Delta \text{F}=0$, when $\Delta \text{F}>0$, and when $\Delta \text{F}<0$.
Then the effectiveness or benefit of each strategy is calculated.
The more beneficial strategy is chosen and it will be passed on to the offspring in an attempt to create a better strategy.


\begin{figure*}[t]
\centering
\begin{tabular}{cccc}
 \includegraphics[width=4.2cm]{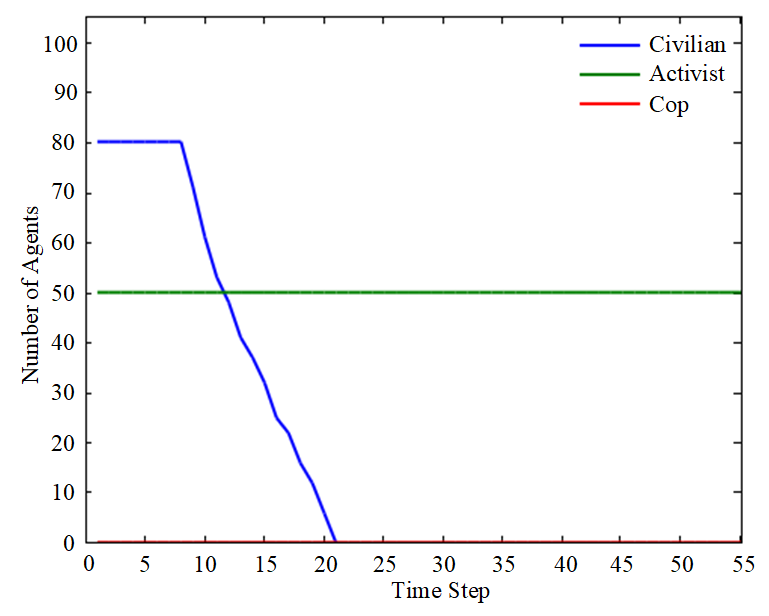} &  \includegraphics[width=4.2cm]{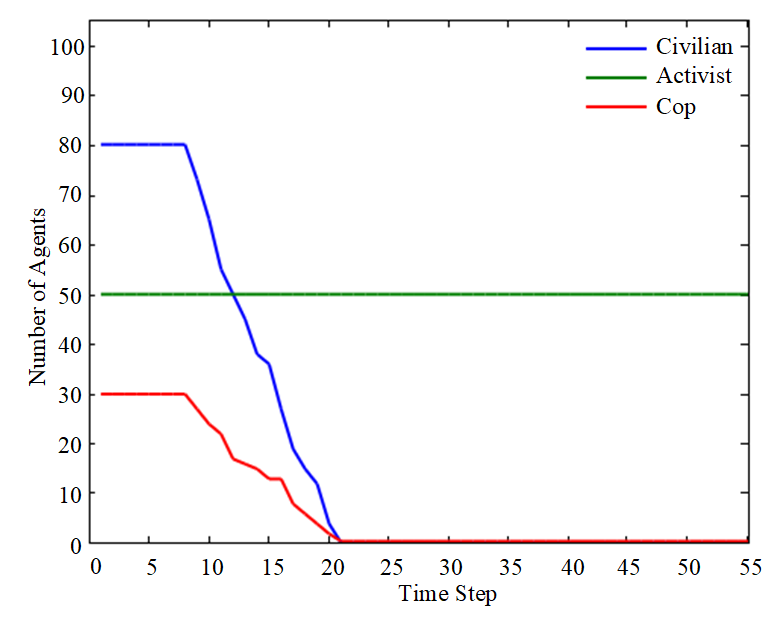} & \includegraphics[width=4.2cm]{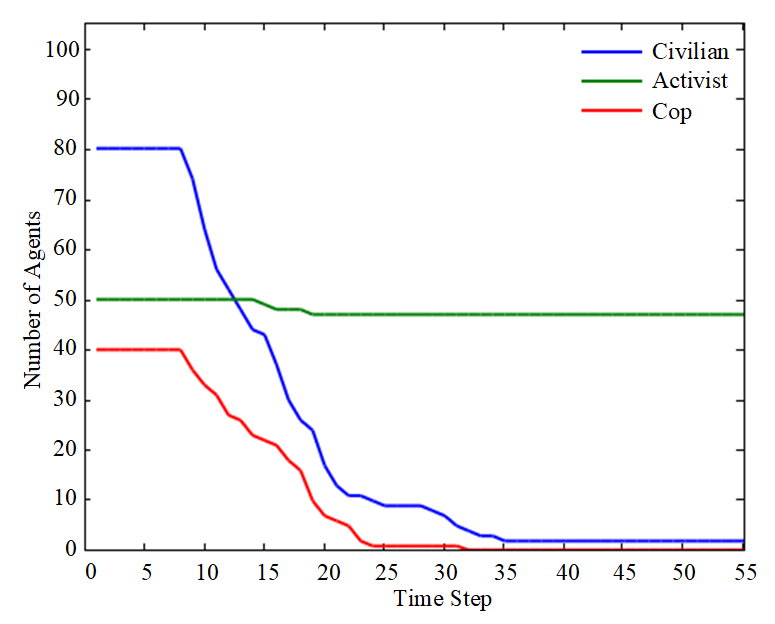} &  \includegraphics[width=4.2cm]{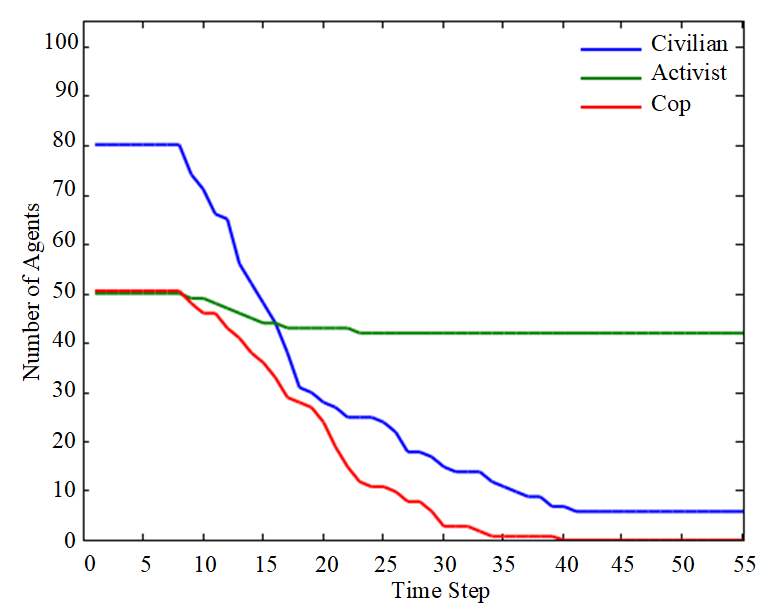}\\
 (a)   & (b)  & (c)  & (d)  \\
   \includegraphics[width=4.2cm]{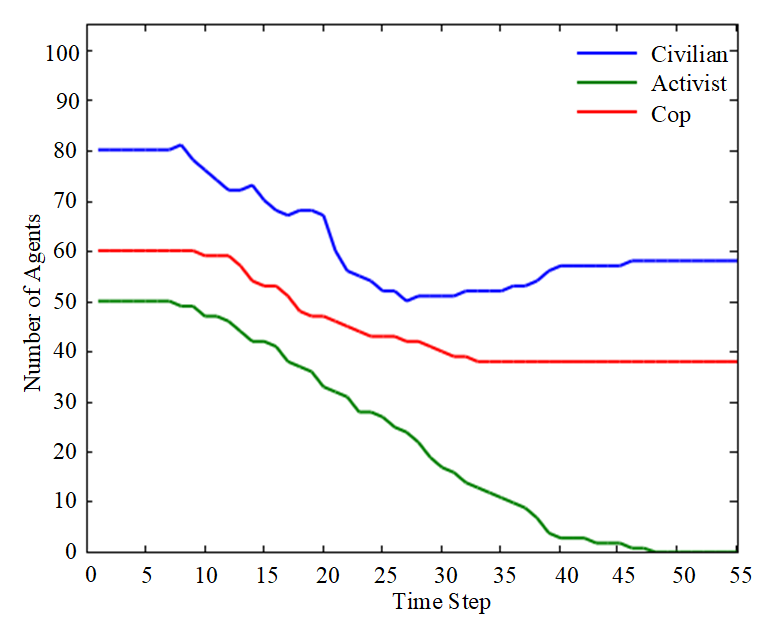} &  \includegraphics[width=4.2cm]{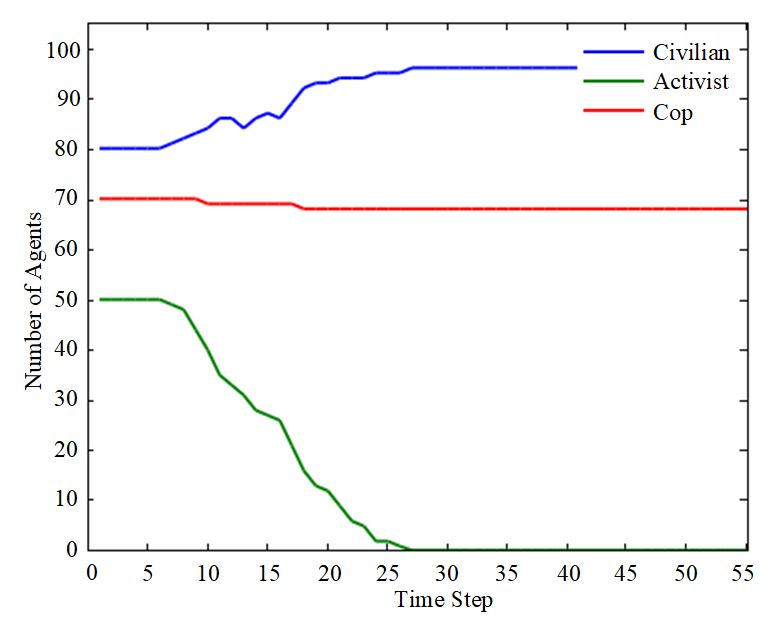} & \includegraphics[width=4.2cm]{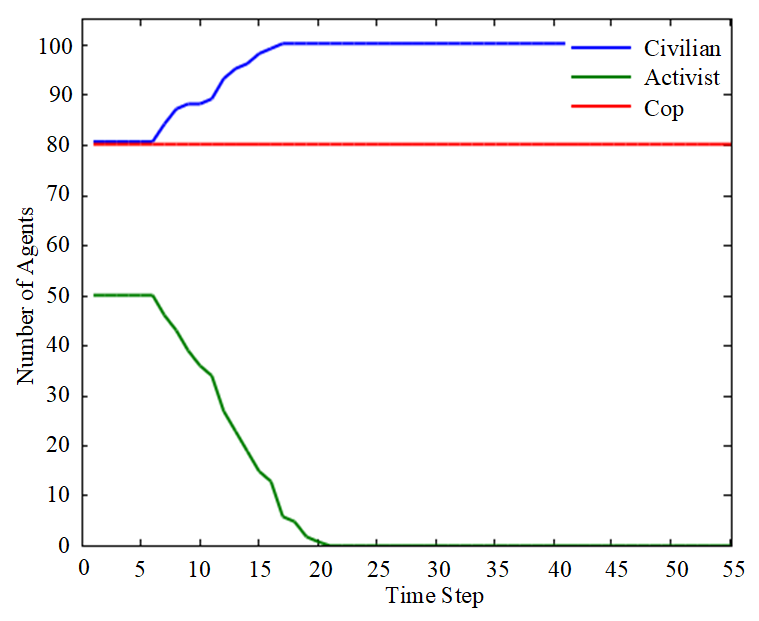} &  \includegraphics[width=4.2cm]{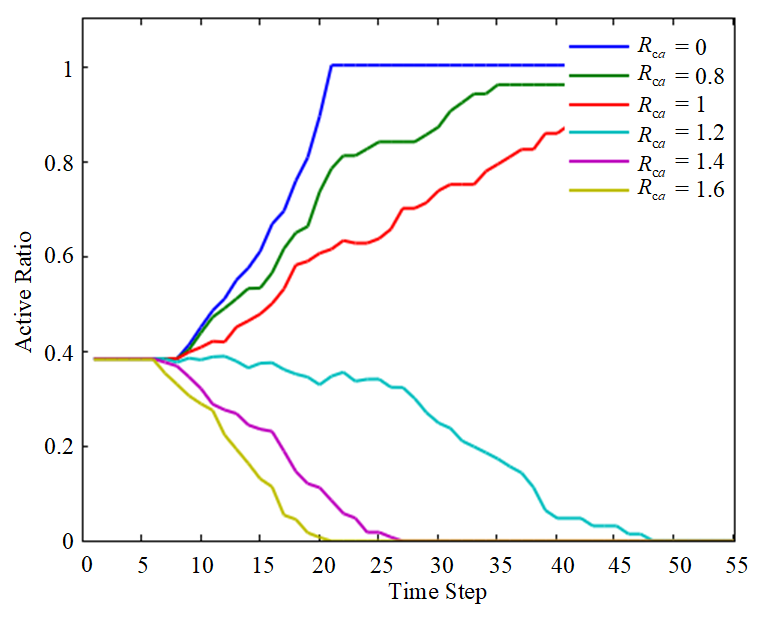}\\
 (e)  & (f)  & (g)  & (h)  \\
\end{tabular}
\caption{
The numbers of three types of agents at different time steps according to different initial $R_{ca}$
(the ratio of cops to activists): (a) $R_{ca}$ is 0 (0 cops),
(b) $R_{ca}$ is 0.6 (30 cops), (c) $R_{ca}$ is 0.8 (40 cops), (d) $R_{ca}$ is 1 (50 cops), (e) $R_{ca}$ is 1.2 (60 cops),
(f) $R_{ca}$ is 1.4 (70 cops), and (g) $R_{ca}$ is 1.6 (80 cops).
(h) Active ratio curves for different initial $R_{ca}s$.
(a) shows a steep drop in the number of civilians. The number of activists remains the same because there are no
cops to fight with the activists. We can see from (a) to (g) that the number of dead activists increases as
$R_{ca}$ increases. When $R_{ca}$ is 1.2, all the activists have been subdued by cops. The time required for all the activists to be
subdued by cops decreases as $R_{ca}$ increases. The civilian survival time lengthens
and there are more surviving civilians as $R_{ca}$ increases. We can see that there is an initial rise
in the number of civilians in (f) and (g) because the total deterrent force of the cops is much higher
than that of the activists. Some activists change their roles (from activist to civilian).
(h) shows the overview of active ratios (ratios of the activists to the sum of the activists and civilians) corresponding to different
$R_{ca}s$. An inverse relationship between the active ratio and the $R_{ca}$ is presented. Therefore, increasing the
ratio of cops can help subdue activists.
}
\label{nc}
\end{figure*}

\subsection{Behavior control module} \label{Behavior mechanism}

The behavior control method of agents is determined by antagonistic emotional contagion and evolutionary game theoretic approaches. In this section, we present some rules about how agents determine their positions at the next time step and their living states.

Agents determine their positions at the next time step based on the cellular automaton model \cite{009}.
A cellular space of $M\times N$ cells is defined and each agent occupies one cell.
At each time step, agents choose to move to their neighboring cells or stay still.

Whether an agent moves or not depends on the deterrent forces exhibited by his or her neighboring agents.
For a cop or an activist, this is divided into the following possible cases according to real-world videos:

\begin{itemize}
\item If the total deterrent force of agents in the opposing group is higher than that of agents of the same type in agent $i$' s
neighboring cells, he or she has to move.
The moving direction of agent $i$ is determined by the expected benefits of his or her neighboring cells.
At first, the neighboring cells around agent $i$ will be checked to find the empty ones (an empty cell means that there is not an agent in it).
Then the expected benefits of all the empty cells are calculated. The cell with the highest benefit is the position to which agent $i$ will move.

\item Next we consider the situation where the total deterrent force of the opposing agents is less than that of agents of the same type
in agent $i$'s neighboring cells.
If agent $i$'s strategy is defection, he or she will move to the nearest opposing agent (i.e. attack
the opposing agent). If agent $i$'s strategy is cooperation, he or she will stay away from opposing
agents and move to the nearest civilian. If agent $i$ is a cop, he or she will protect the civilian. If agent $i$ is an
activist, he or she will attack the civilian. In this case, agent $i$ may also choose to stay still.

\item Agent $i$ with no neighbors chooses to move. The moving direction is the same as the situation where the total deterrent force of the opposing agents is less than that of agents of the same type.


\item Civilians move to safer positions where there are more cops around them.
\end{itemize}

The agent with a defection strategy attacks his or her opponents.
The agent may be dead. A dead agent in this case means being subdued by his or her opponents and therefore posing no threat to these opponents.
It doesn't mean real death. At each time step,
the death probability of each agent is calculated and denoted as ${{P}_{die}}$. In contrast to the definition in \cite{027}, which is based on the number of cops and activists, we define
${{P}_{die}}$ based on the total deterrent forces of cops and activists.

%
\begin{equation}\label{eq17}
{{P}_{die}}=1-\exp \left( ln0.1 \cdot \frac{\sum{{{F}_{i}}}}{{{\sum{{\overset{\scriptscriptstyle\frown}{F}}}}_{i}}} \right)
\end{equation} 
where $\sum {{F}_{i}}$ represents the total deterrent force of agents of the same type, and $\sum \widehat{{{F}_{i}}}$
represents the total deterrent force of his or her opponents in the cells neighboring agent $i$.
$\frac{\sum{{{F}_{i}}}}{{{\sum{{\overset{\scriptscriptstyle\frown}{F}}}}_{i}}}$ denotes the total deterrent force of the cop-to-activist ratio within the perceived range of agent $i$. $ln0.1$ is set to ensure a plausible value (${P}_{die}=0.9$) when ${\sum{{{F}_{i}}}}={{{\sum{{\overset{\scriptscriptstyle\frown}{F}}}}_{i}}}$ \cite{117,027}.

Each agent has an early warning threshold $T_{warn}$. When the value of ${{P}_{die}}$ exceeds
the threshold $T_{warn}$, the value of $warn\_time$ increases by 1. When the value
of $warn\_time$ exceeds the threshold $T_{warn\_time}$, the agent will die.
Because the endurance of each agent is different, the values of the thresholds
$T_{warn}$ and $T_{warn\_time}$ are also different for each agent.

\section{Implementation and performance}
\label{experiments}

We have implemented our model using Visual C++ to simulate antagonistic crowd behaviors based on Unity3D .
The computing environment is a common PC with a quadcore 2.50 GHz CPU,16 GB memory, and an Nvidia GeForce GTX 1080 Ti graphics card.

\begin{table}[htbp]\tiny
\setlength{\belowcaptionskip}{10pt}
\caption{List of parameter values used in our simulation}
\setlength{\tabcolsep}{0.6mm}{
\begin{tabular}{|l|c|c|c|c|c|c|c|c|c|}
\hline
\multicolumn{1}{|c|}{\multirow{2}{*}{Scenario}}       & \multicolumn{3}{c|}{Number of agents} & \multirow{2}{*}{Size of 2-D Grid} & \multirow{2}{*}{$T_{a2c}$} & \multirow{2}{*}{$T_{c2a}$} & \multicolumn{3}{c|}{Emotion}                                                   \\ \cline{2-4} \cline{8-10}
\multicolumn{1}{|c|}{}                                & Civilians    & Activists    & Cops    &                                   &                             &                             & Civilians & Activists                                                   & Cops \\ \hline
\multicolumn{1}{|c|}{\begin{tabular}[c]{@{}c@{}}No.1:Activists attack \\civilians\end{tabular}} & 80           & 50           & 40      & 20$\times$20 squares                     & 0.1                         & -0.5                        & 0.1       & -0.5                                                        & 0.5  \\ \hline
No.2:Role transitions                                 & 80           & 50           & 70      & 20$\times$20 squares                     & 0.1                         & -0.5                        & 0.1       & -0.5                                                        & 0.5  \\ \hline
\begin{tabular}[c]{@{}c@{}}No.3:Cops encircling\\ activists\end{tabular}                        & 80           & 50           & 70      & 20$\times$20 squares                     & 0.1                         & -0.5                        & 0.1       & -0.5                                                        & 0.5  \\ \hline
No.4:Real-world 1                                     & 10           & 50           & 30      & 20$\times$20 squares                     & 0.1                         & -0.5                        & 0.1       & -0.5                                                        & 0.5  \\ \hline
No.5:Real-world 2                                     & 80           & 50           & 30      & 20$\times$20 squares                     & 0.1                         & -0.5                        & 0.1       & -0.5                                                        & 0.5  \\ \hline
No.6:Real-world 3                                     & 3            & 14           & 40      & 20$\times$20 squares                     & 1                           & -1                          & 0.1       & \begin{tabular}[c]{@{}c@{}}-0.1 \\ and \\ -0.3\end{tabular} & 0.9  \\ \hline
No.7:Real-world 4                                     & 0            & 30           & 100     & 40$\times$40 squares                     & 1                           & -1                          & 0         & -0.2                                                        & 0.8  \\ \hline
No.8:Real-world 5                                     & 100          & 30           & 100     & 40$\times$40 squares                     & 1                           & -1                          & 0.1       & \begin{tabular}[c]{@{}c@{}}-0.9\\ and \\ -0.2\end{tabular}  & 0.9  \\ \hline
No.9:ACSEE vs. CVM                                    & 80           & 50           & 30      & 20$\times$20 squares                     & 0.1                         & -0.5                        & 0.1       & -0.5                                                        & 0.5  \\ \hline
\end{tabular}}
\label{pv}
\end{table}

We run a series of experiments involving varying role number ratios in outdoor  scenarios.
The parameter values in different scenarios used in the simulation runs are listed in Table \ref{pv}.
The perceived range \textit{PR} of agents is 10, $T_{warn} \in [0.7,0.9]$, and $T_{warn\_time} \in [8,20]$.
Figures \ref{nc}, \ref{fig:ar_pr}, \ref{fig:ar_a}, \ref{fig:22}, and \ref{fig:252627} report the results of 200 runs. The average of the 200 runs is the final result.
We investigate the effect of varying role number ratios on the crowd violence results (in Section \ref{The impact of the initial number of cops on the final result of group violence}).
 By analyzing the simulation results, we find that our model can account for many emergent phenomena (discussed in Section \ref{Some fascinating emergent phenomena our model uncover}).
Next, we study how different important factors influence crowd violence.
In Section \ref{The impact of emotional contagion on the final result of group violence} we show the impact of emotional contagion on the results of crowd violence.
We compare the different simulation results with or without considering emotional contagion. In Section \ref{The influence of deterrent force on strategy}, the relationship between the deterrent force and the strategy is revealed.
The strategies adopted by each agent in different situations are analyzed.
In Section \ref{Comparisons}, our simulation results are compared with the real-world videos and the simulation results generated by different models. The real-world videos are chosen from the public dataset and real antagonistic events. More details about comparisons can be seen in the supplementary video. User studies are performed to evaluate our method in Section \ref{User Studies}.


\subsection{The impact of ratios of agents in different roles on the result of crowd violence}
\label{The impact of the initial number of cops on the final result of group violence}

\begin{figure}[t] \begin{centering}
  \centering
  \includegraphics[width=7.5cm]{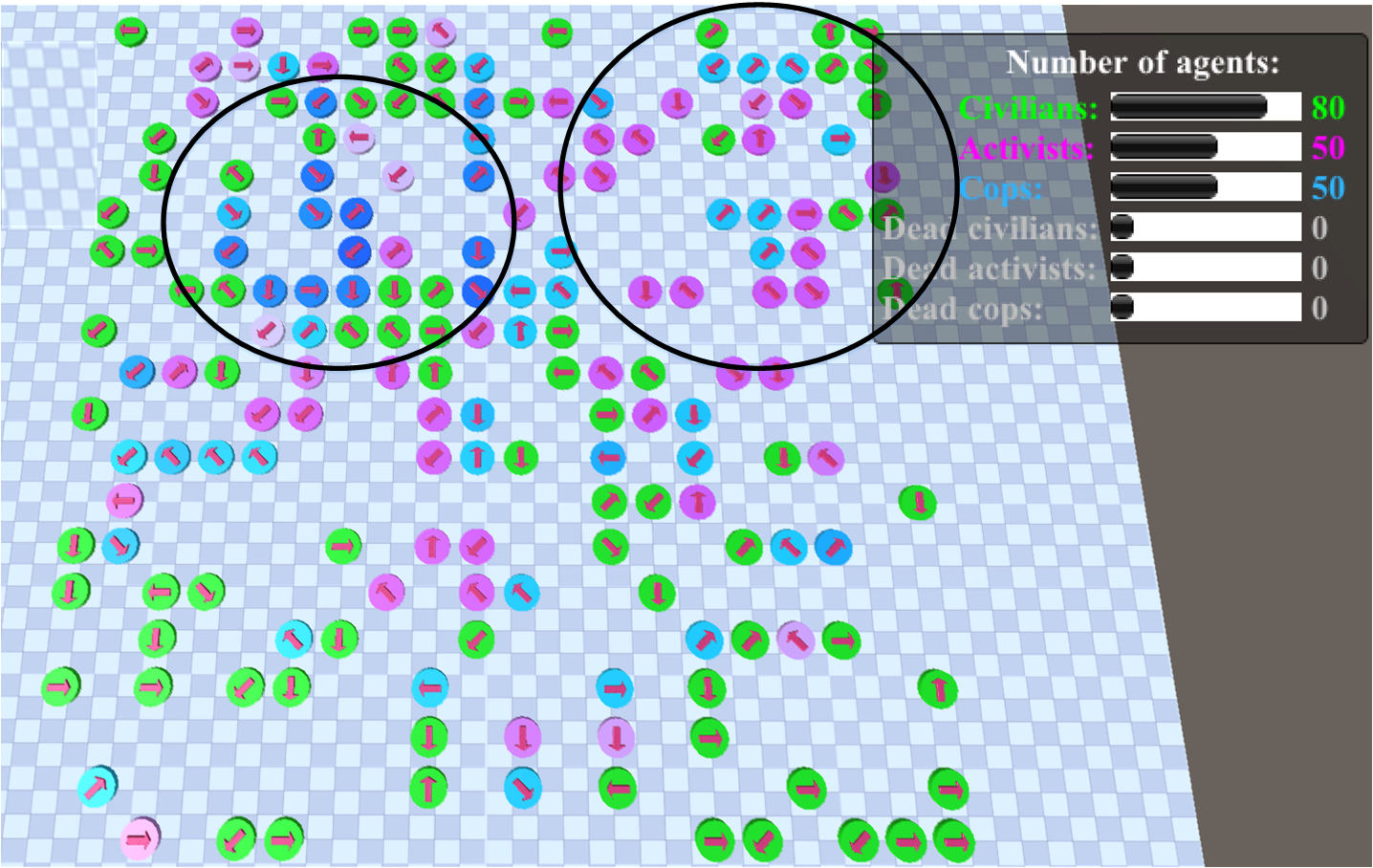}
  \centering
  \caption{The positions of all the agents at the 168th frame ($R_{ca}$ is 1).
 The green, purple, and blue circles represent civilians, activists, and cops, respectively.
 The stronger the color intensity of a circle, the higher the deterrent force of the agent.}
  \label{fig:18}
  \end{centering}
\end{figure}

\begin{figure*}[t]
\centering
\begin{tabular}{ccc}
 \includegraphics[width=5.5cm]{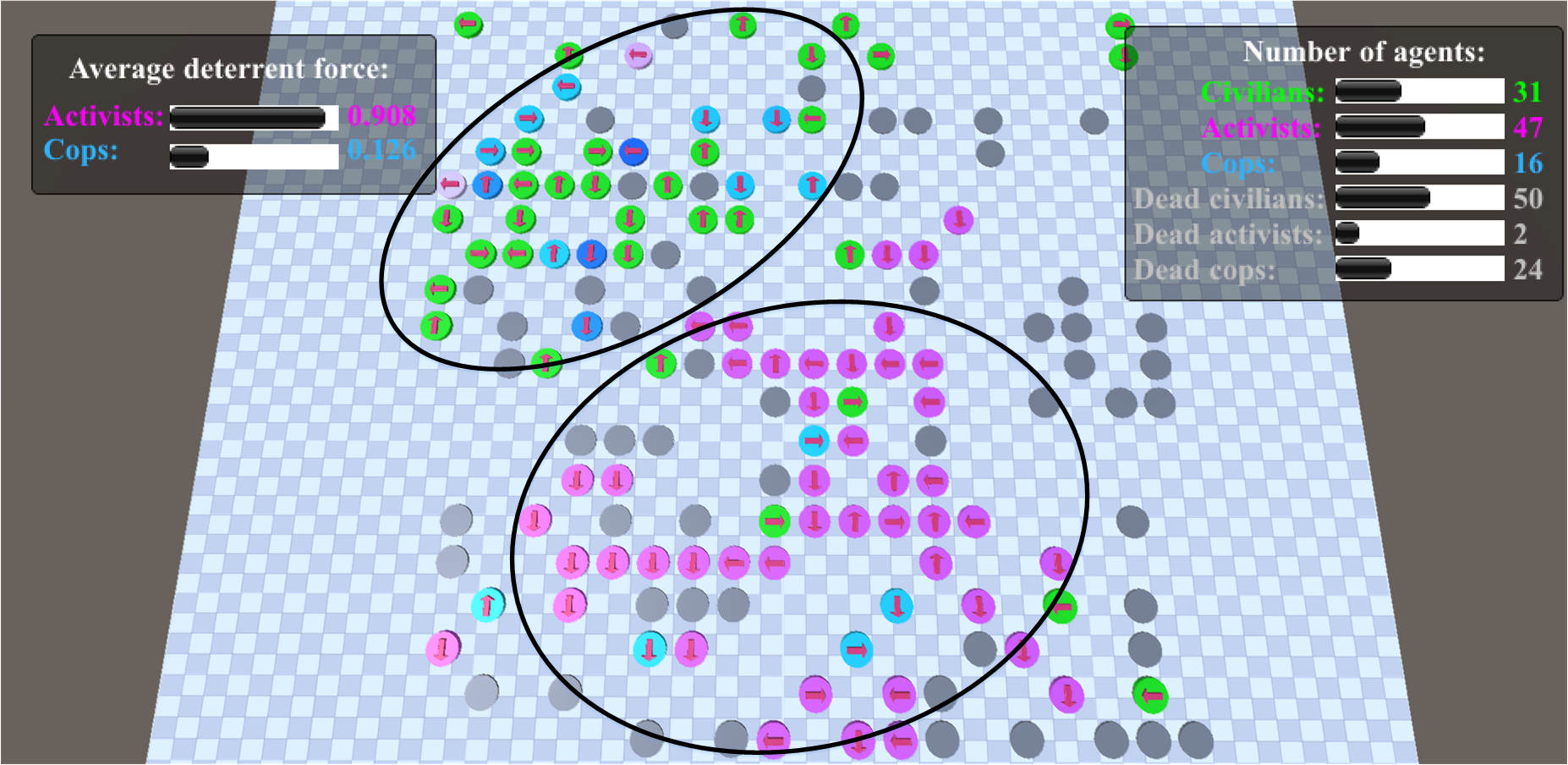} &  \includegraphics[width=5.5cm]{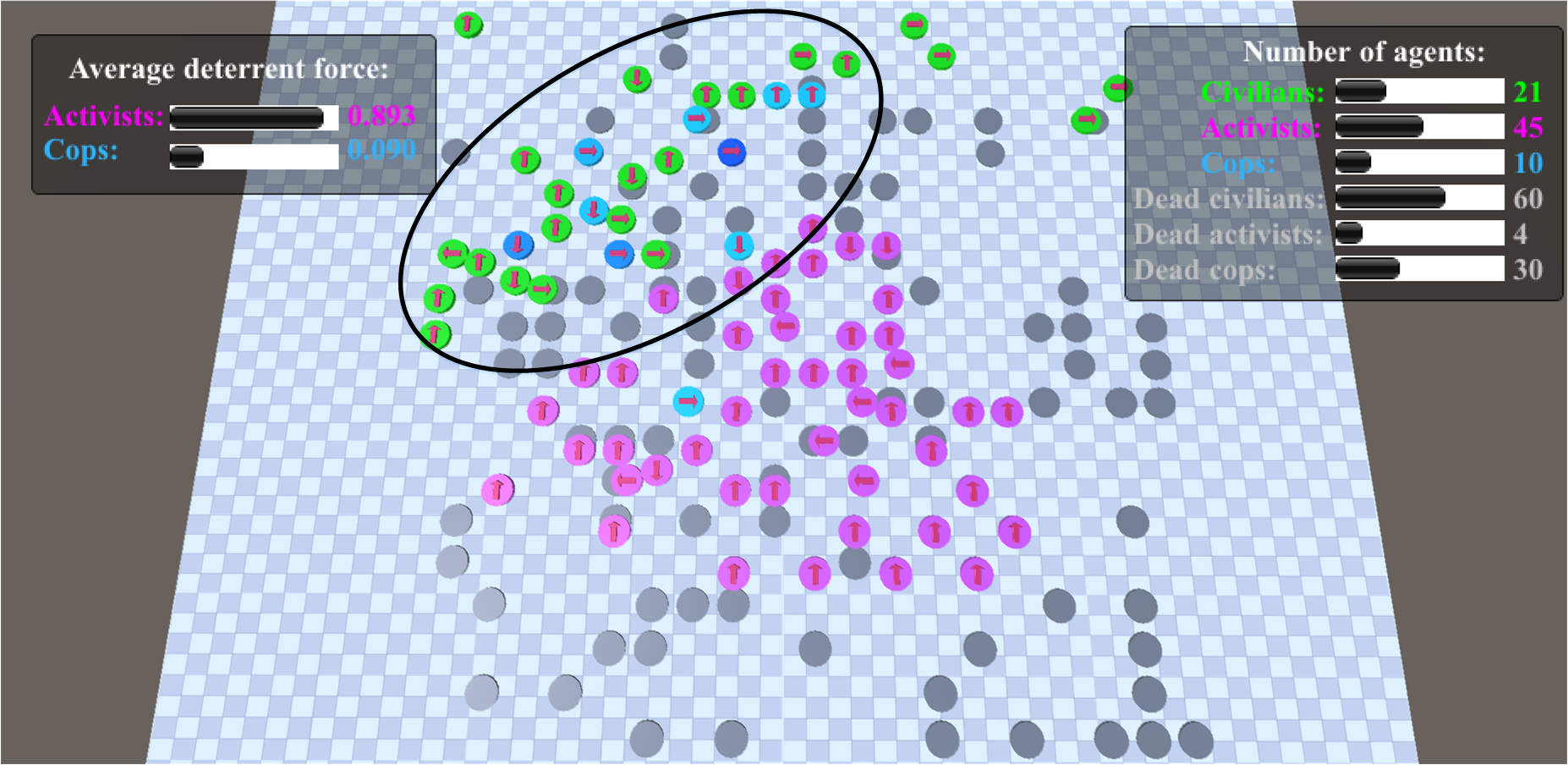} & \includegraphics[width=5.5cm]{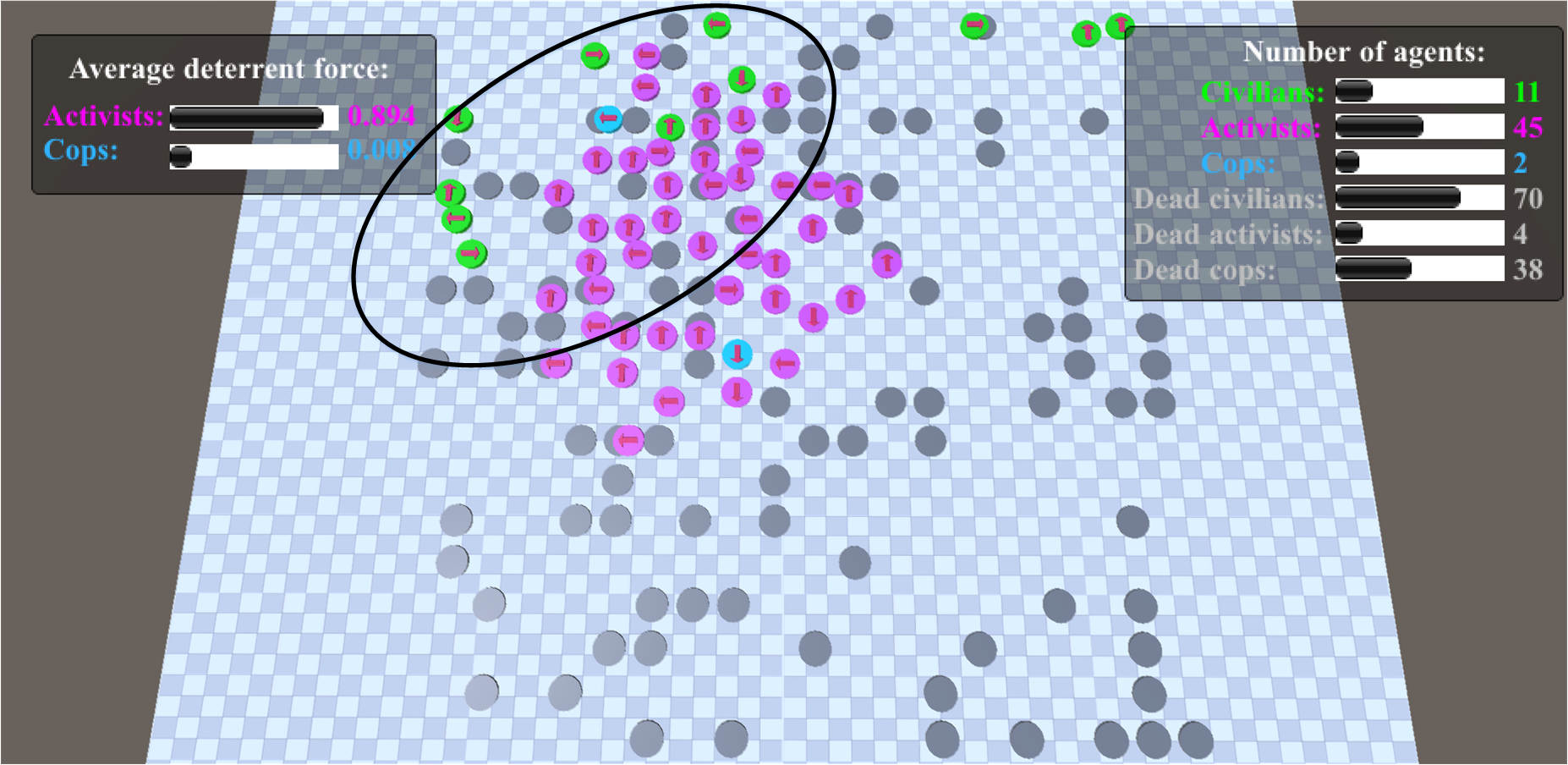}\\
 (a)   & (b)  & (c)   \\
  \includegraphics[width=5.5cm]{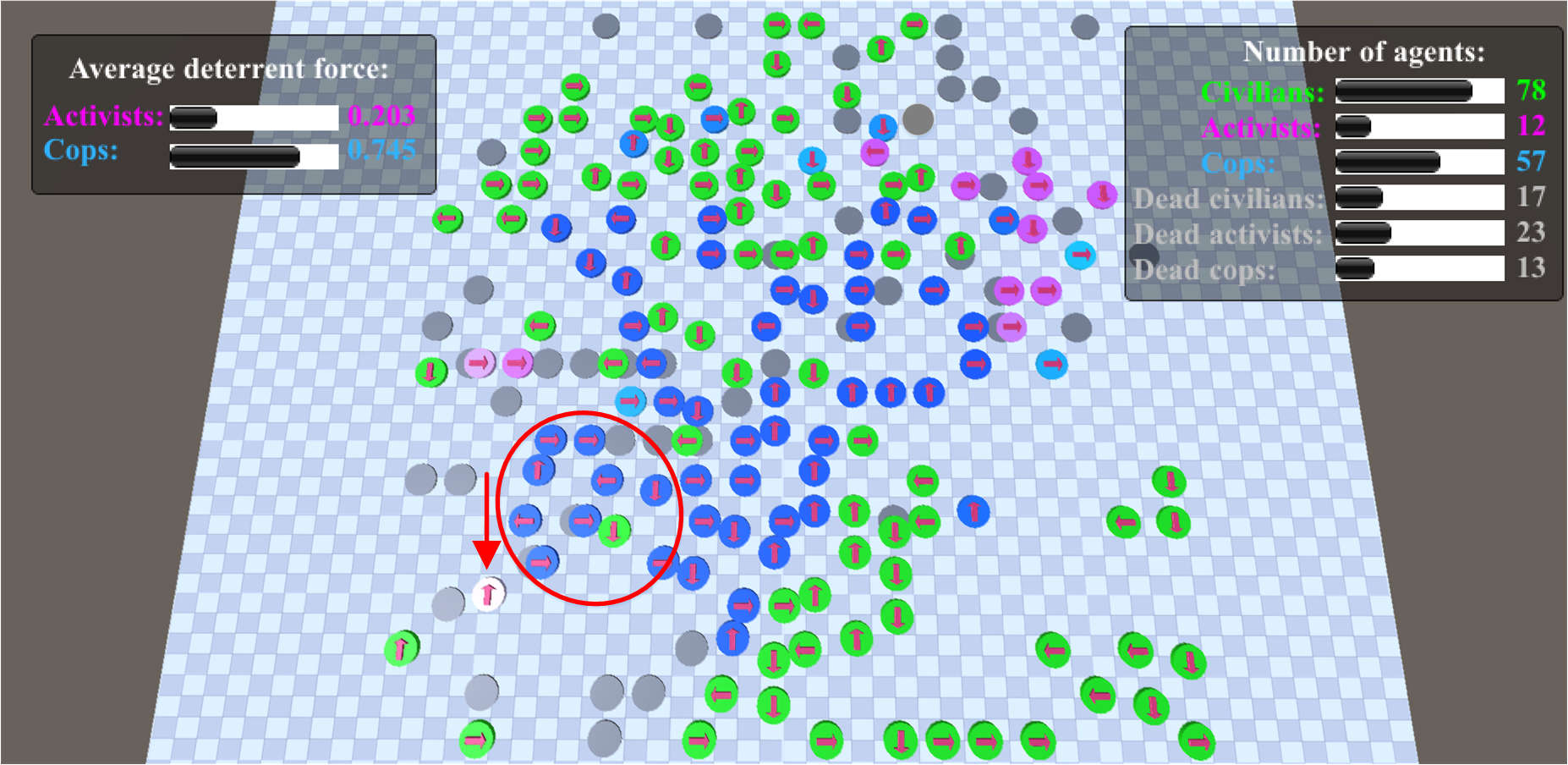} &  \includegraphics[width=5.5cm]{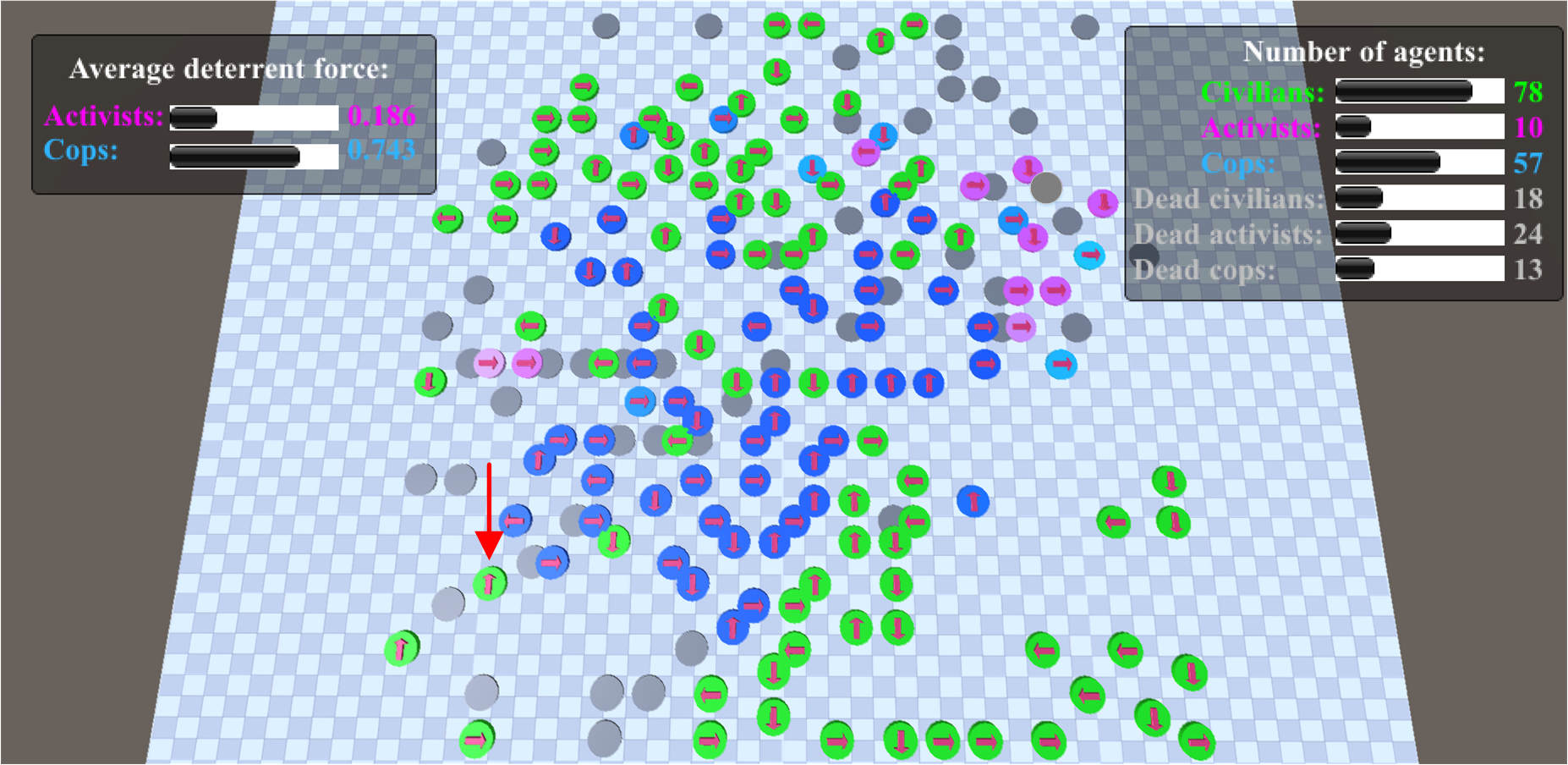}&  \includegraphics[width=5.5cm]{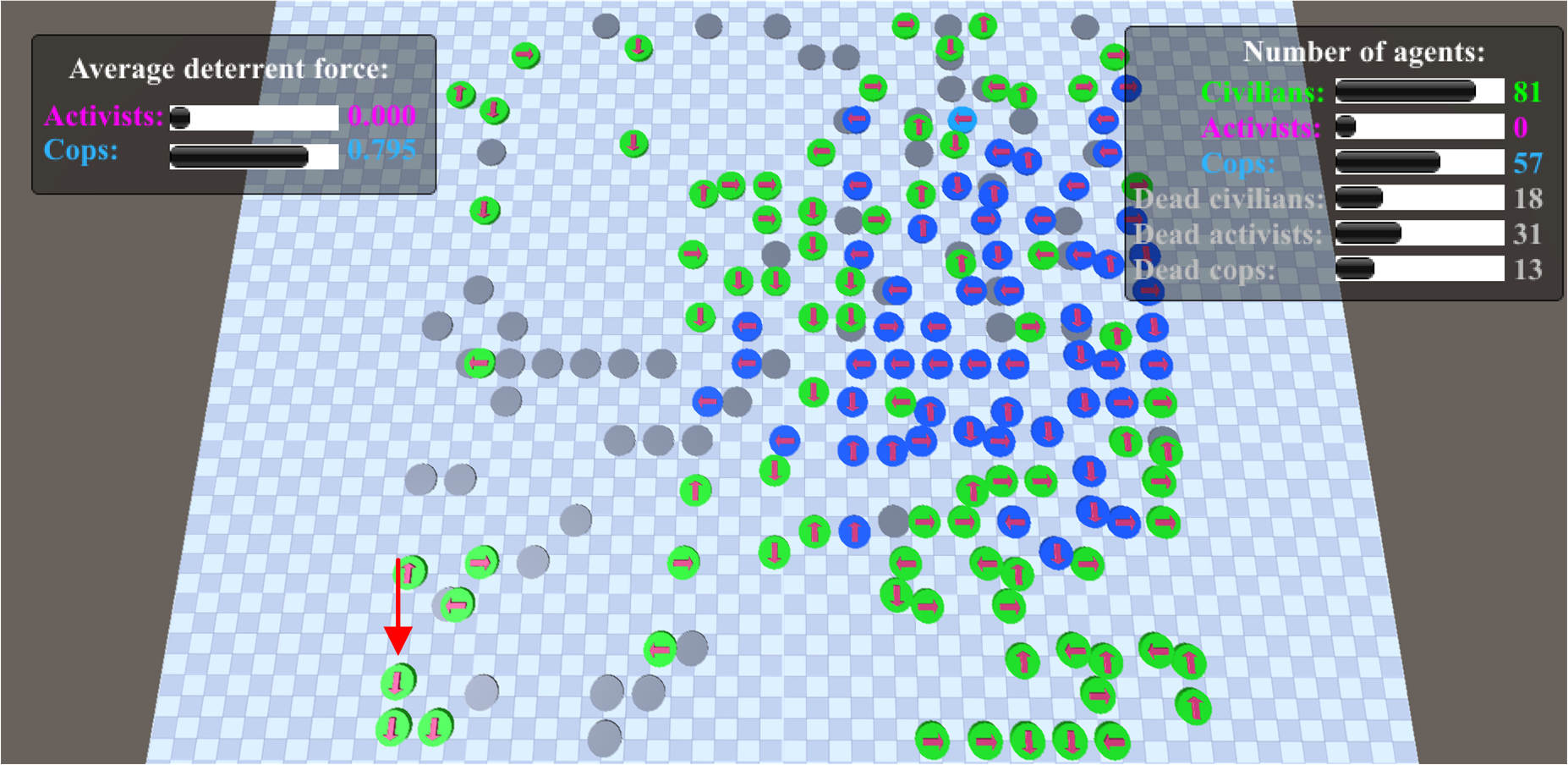}\\
 (d)   & (e)  &(f)  \\
  \includegraphics[width=5.5cm]{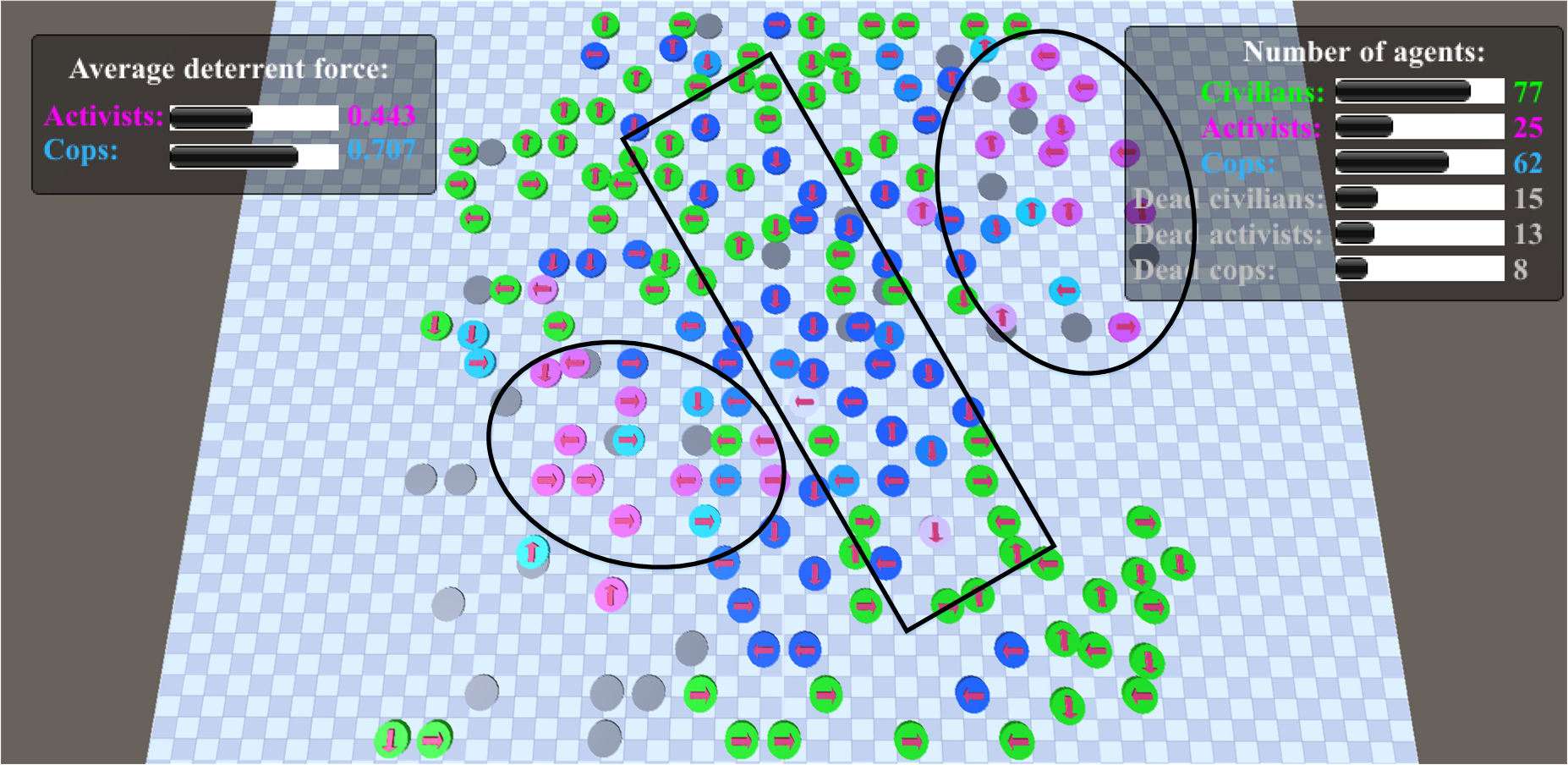} &  \includegraphics[width=5.5cm]{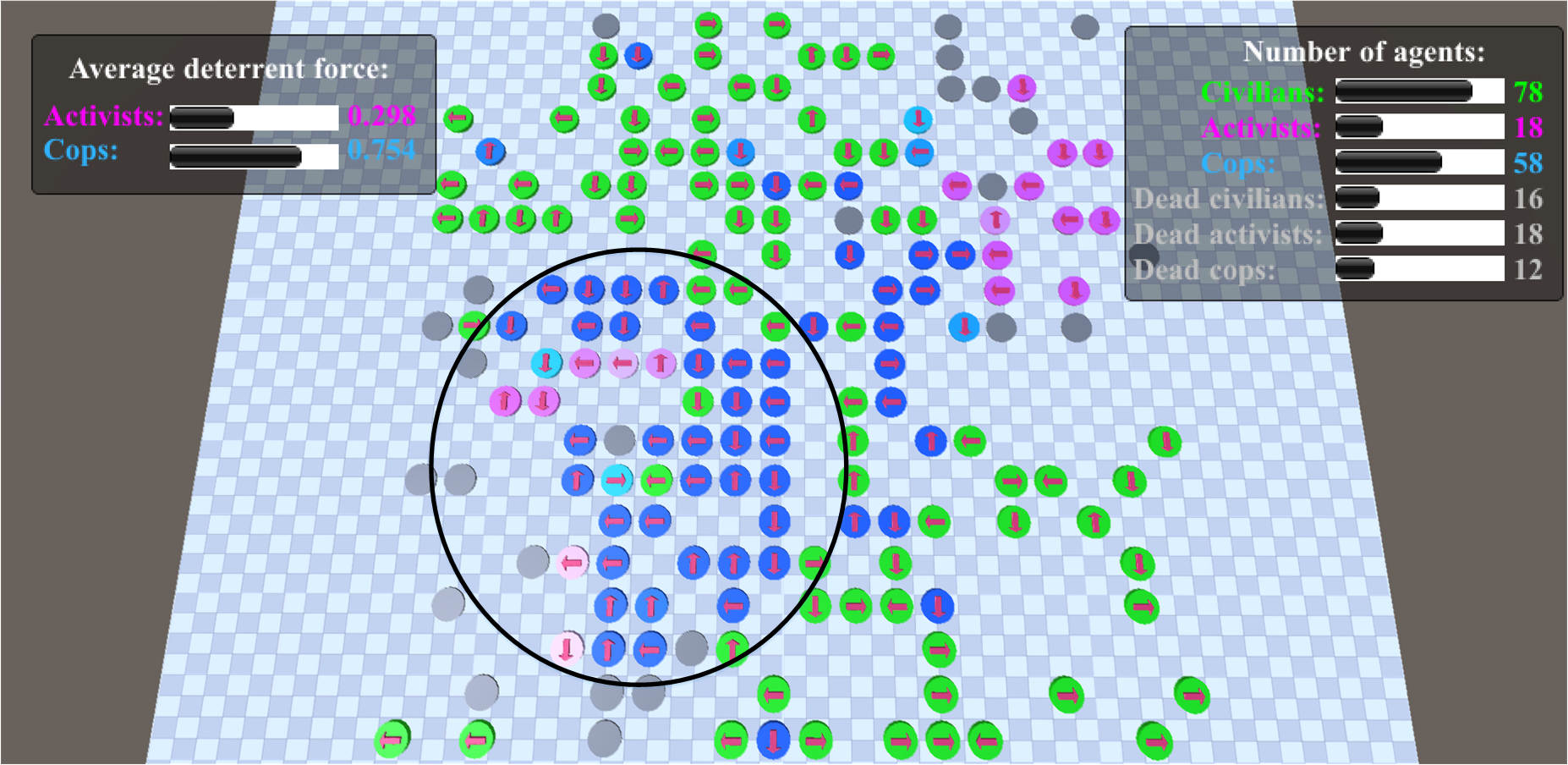} & \includegraphics[width=5.5cm]{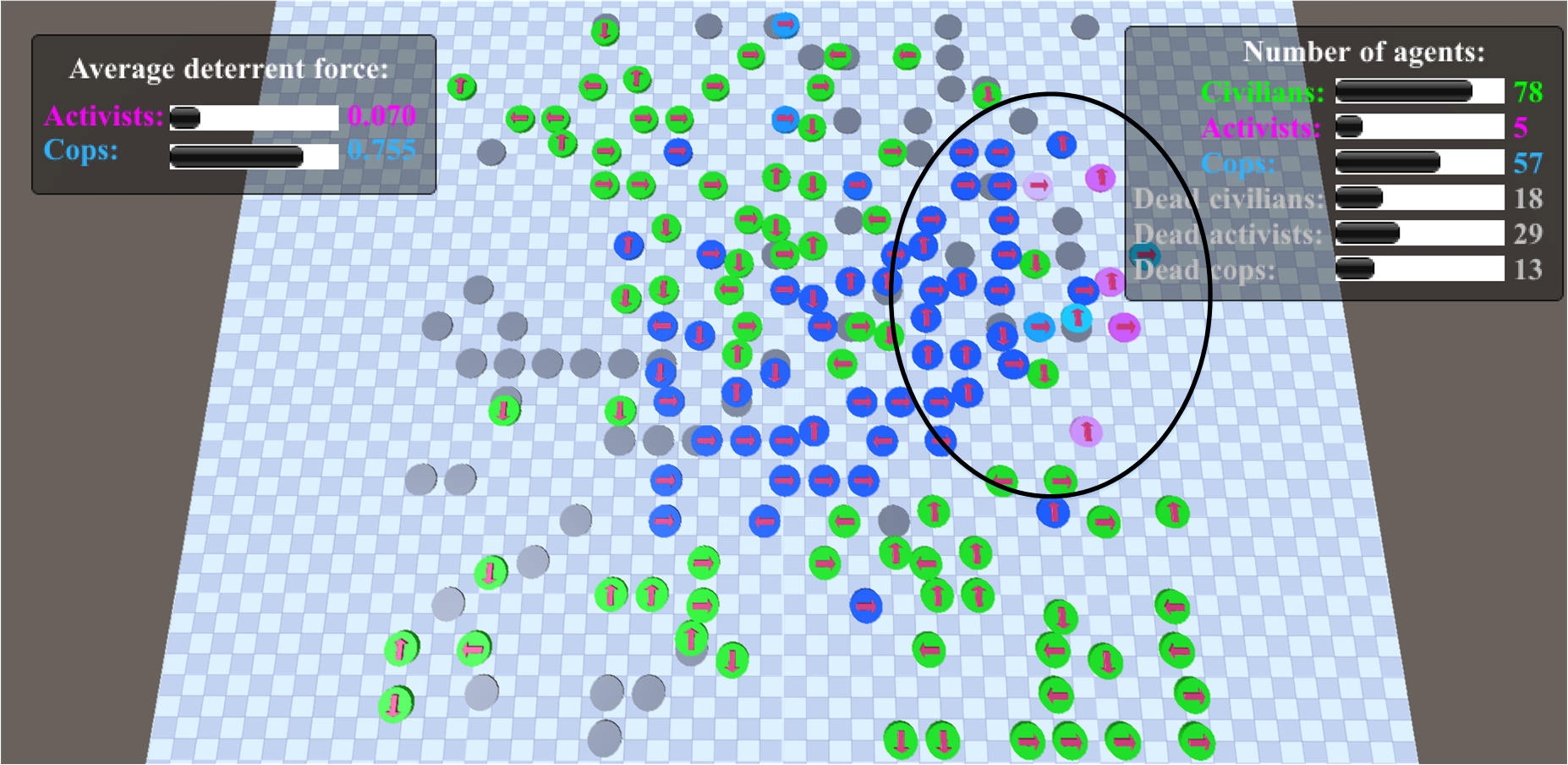}\\
 (g)   & (h)  & (i)   \\
\end{tabular}
\caption{
Some fascinating emergent phenomena our model uncovers.
(a), (b), and (c) are the simulations of activists attacking civilians.
(d), (e), and (f) are the simulations of role transitions.
(g), (h), and (i) are the simulations of cops encircling activists.
The green, purple, blue, and grey circles are civilians, activists, cops, and dead agents, respectively.
The stronger the color intensity of a circle, the higher the deterrent force of the agent.
}
\label{fig:19}
\end{figure*}

We investigate the effects of varying ratios of agents in different roles on the results of crowd violence.
By analyzing a large number of real-world antagonistic videos, we select several representative values of $R_{ca}$ (the ratio of cops to activists). In this section the initial $R_{ca}$s are 0, 0.6, 0.8, 1, 1.2, 1.4, and 1.6.
The initial numbers of civilians and activists are 80 and 50, respectively.
The initial emotion values of all the activists and cops are -0.5 and 0.5, respectively.

%

We can learn from Figure \ref{nc} that the number of different kinds of agents determines the outcome of crowd violence to some extent.
When we increase the ratio of cops, they can subdue activists quickly, stabilize the situation, and reduce civilian casualties.
When $R_{ca}$ is 1, it means that the initial
total deterrent force of the cops is equal to that of the activists. Both cops and activists have the same probability of winning. However, in the end, the cops fail which means that all the cops are subdued by the activists.
We offer a detailed explanation in the following analysis.

Figure \ref{fig:18} shows the positions of all the agents at the 168th frame when $R_{ca}$ is 1. At this time step, the numbers of cops and activists are the same.
Activists with strong deterrent forces are mainly located in the upper right corner
of the scene. Some cops with weak deterrent forces are among the crowd of activists
(upper right corner of the scene). Some surrounding civilians turn into activists because of the incitement of the activists with high deterrent forces. The cops with strong deterrent forces are in the upper left corner of the scene, which makes it difficult for them to rescue the cops in the upper right corner.
When the cops with weak deterrent forces are killed by the activists with strong deterrent forces, there are fewer cops left.
Finally, all the activists get together to attack the remaining cops and the activists win.

We learn from Figure \ref{fig:18} that activists participate in collective behavior to create regions of low cop-to-activist ratios, which reduces the chances of activist death.
The conglomeration of scattered activists into small groups and the amalgamation of small groups into large ones make it difficult to wipe them out \cite{117}. Although the number of agents determines the outcome of crowd violence to a certain extent,
 some take advantage of agents' spatial distributions to affect the outcome of crowd violence.

\subsection{The impact of agent parameters on the result of crowd violence}
\label{Analysis of agent parameters}

In this section we analyze the impact of agent parameters on the result of crowd violence. We choose two parameters which have great influences on the simulation results: \textit{PR} (the radius of perceived range) and $A$ or $B$ in Equation \ref{eq2}. In Equation \ref{eq2}, $A_{j,i}$ is the strength attribute by which an emotion is received by $i$ from sender $j$ and $B_{i,j}$ is the strength attribute by which an emotion is sent from $i$ to receiver $j$.
We discuss the relationship between active ratios (ratios of activists to the sum of the activists and civilians) and the parameters.
The initial numbers of the civilians, activists, and cops are 80, 50, and 70, respectively. The initial emotions of the civilians, activists, and cops are 0.1, -0.5, and 0.5, respectively. The initial total deterrent force of the cops is higher than that of the activists.

\begin{figure}[t] \begin{centering}
  \centering
  \includegraphics[width=7.5cm]{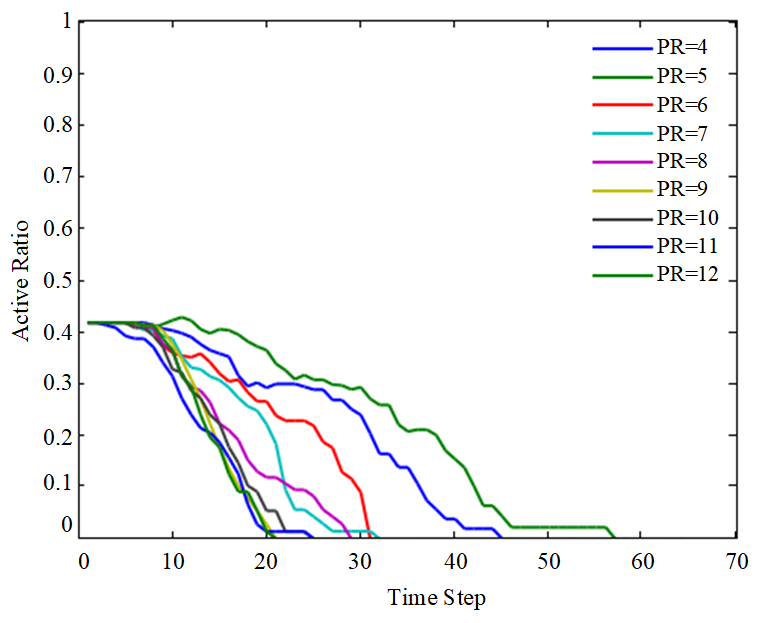}
  \centering
  \caption{ Active ratios corresponding to different \textit{PRs} (the radius of perceived range). The active ratio decreases with the increase of \textit{PR}. When \textit{PR} is greater than 10, the active ratio tends to be stable. We choose $\textit{PR}=10$ in this paper. }
  \label{fig:ar_pr}
  \end{centering}
\end{figure}

We show active ratios according to various values of \textit{PR} in Figure \ref{fig:ar_pr}. We assume that all individuals have the same \textit{PR}. When \textit{PR} of agents are different, the results of active ratios are also different. With
the increase of \textit{PR}, agents can more accurately estimate the situations. Therefore, the high total deterrent force of
cops plays a role in defeating the activists. The active ratio decreases with the increase of \textit{PR}.

\begin{figure}[t] \begin{centering}
  \centering
  \includegraphics[width=7.5cm]{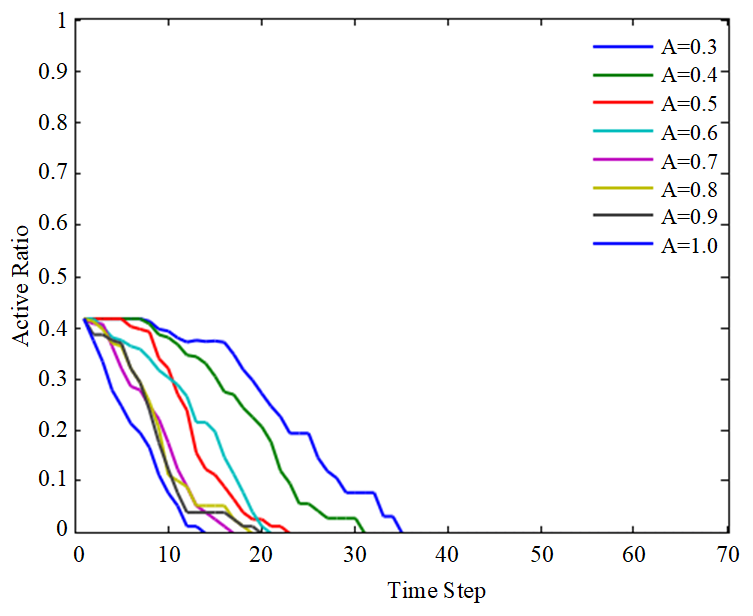}
  \centering
  \caption{ Active ratios corresponding to different values of $A$. The active rate decreases with the increases of $A$. When $A$ is greater than 0.8, the active ratio tends to be stable. We choose $A=0.8$ in this paper.}
  \label{fig:ar_a}
  \end{centering}
\end{figure}

In addition, we also discuss the relationship between the values of $A$ or $B$ and active ratios in Figure \ref{fig:ar_a}. $A$ and $B$ are very important parameters for emotional contagion in Equation \ref{eq2}. The values of them are positively correlated with the values of emotional contagion.
We suppose all the strength attributes of $A$ between any two individuals are the same and those of $B$ between any two individuals are the same.
 The relationship between the active ratio and $A$ is the same as that between the active ratio and $B$. We take $A$ as an example to discuss this relationship. As the value of $A$ increases, agents' emotions also increase and the difference between the total deterrent forces of cops and
activists is greater. Due to the high deterrent force of cops, more and more activists are subdued by cops. Therefore when the
value of $A$ increases, the active rate decreases.

\subsection{Emergent phenomena uncovered by our model}
\label{Some fascinating emergent phenomena our model uncover}

Our model can simulate many emergent phenomena that conform to real-world scenes.
Figures \ref{fig:19}a, \ref{fig:19}b, and \ref{fig:19}c  show that activists with strong
deterrent forces attack civilians. At the top left corner of this scene, there are a lot of civilians. Many activists are at the lower right corner
of Figure \ref{fig:19}a. Agents of the same type gather together according to \cite{031}.
Therefore, it is reasonable that agents of the same type gather together. The number of activists is larger than that of the cops.
We can see from Figure \ref{fig:19}a that the total deterrent
force of the activists is stronger than that of the cops. Activists attack civilians (Figure \ref{fig:19}b) and more and more
civilians die (Figures \ref{fig:19}b and \ref{fig:19}c).

At the lower right corner of Figure \ref{fig:19}d, there are plenty of cops with high deterrent forces
near the highlighted activist and his deterrent force is weak.
If he continues to resist, he will die. He therefore transitions roles (from activist to civilian). When all the other activists die,
he survives (Figure \ref{fig:19}f).
When the number of surrounding activists is large enough, it may impel civilians to become activists \cite{095}.




In Figure \ref{fig:19}g, there are many more cops than activists. The total deterrent force of the cops is
much stronger than that of the activists. At first, the cops divide the activists into two groups (Figure \ref{fig:19}g) and
prevent the activists from gathering together to form a larger group. Next, the cops
eliminate these two groups of activists individually. There are some activists in the left side of the scene (the first group).
Cops encircle these activists and more activists will die. When all the activists on the left side are eliminated,
the cops return to the activists on the right side of the scene (the second group). These activists are encircled by the cops.
Finally, all the activists are killed by the cops and the cops win.

\begin{figure}[t] \begin{centering}
  \centering
  \includegraphics[width=8.5cm]{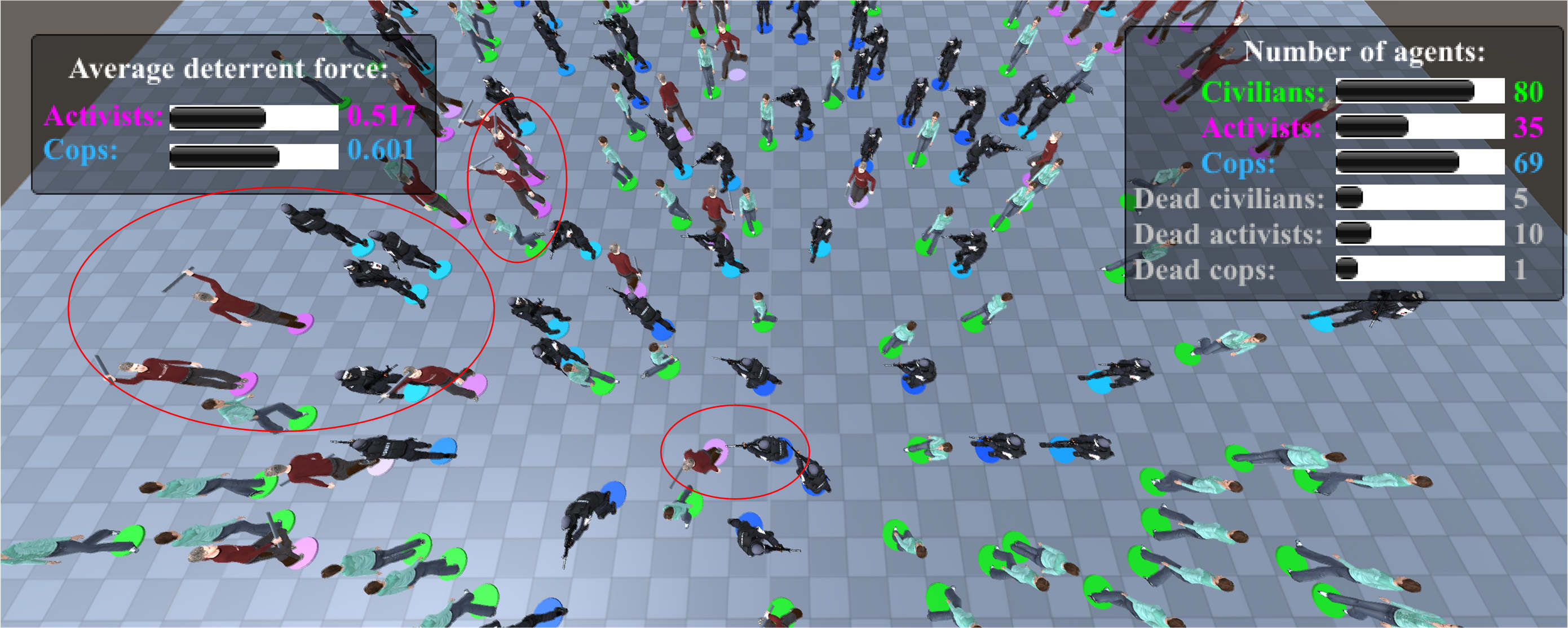}
  \centering
  \caption{ Simulation of antagonistic crowd behavior using 3D character models.}
  \label{fig:2526}
  \end{centering}
\end{figure}

Figure \ref{fig:2526} shows the simulation of antagonistic crowd behavior using 3D character models.
 The activists with high deterrent forces on the left of the scene are not afraid of the cops with low deterrent forces.
 In the upper left corner of the scene, the activists with high deterrent forces attack a civilian. The civilian runs away from the activists.
In the middle and lower part of the scene, the cop with a high deterrent force attacks the activist and he runs away from the cop.


\begin{figure}[t]
\centering
\begin{tabular}{cc}
 \includegraphics[width=4cm]{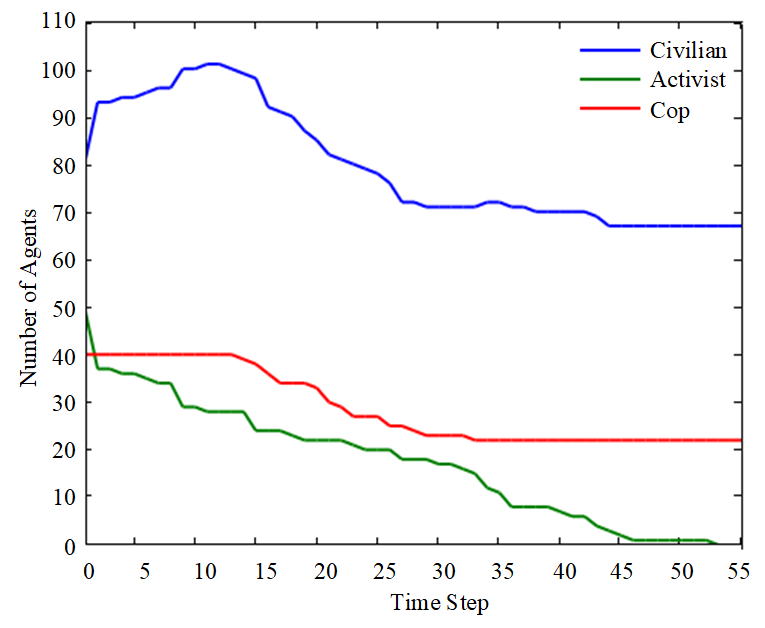} &  \includegraphics[width=4cm]{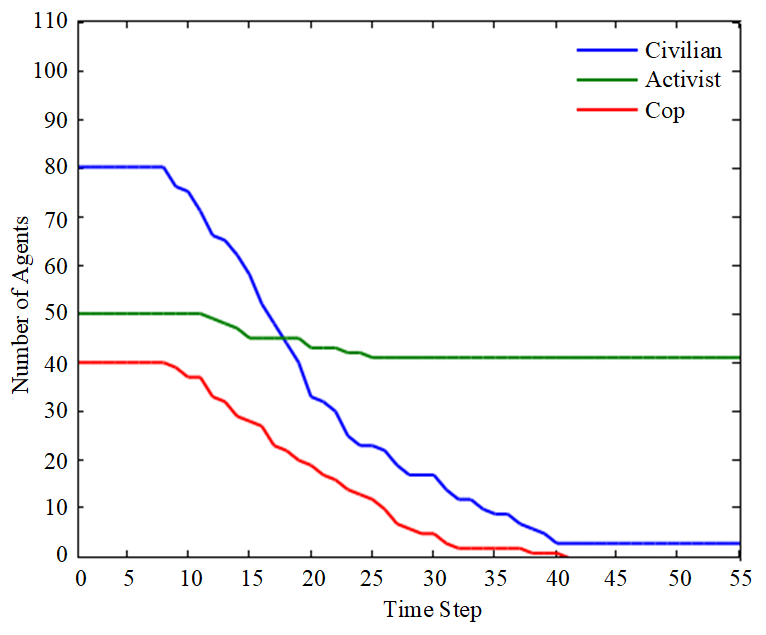} \\
 (a)   & (b)     \\
\end{tabular}
\caption{
The changes in the number of civilians, activists, and cops: (a) considering emotional contagion and (b) not
considering emotional contagion.
}
\label{fig:22}
\end{figure}

\begin{figure}[t]
\centering
\begin{tabular}{cc}
 \includegraphics[width=3.5cm,height=3.5cm]{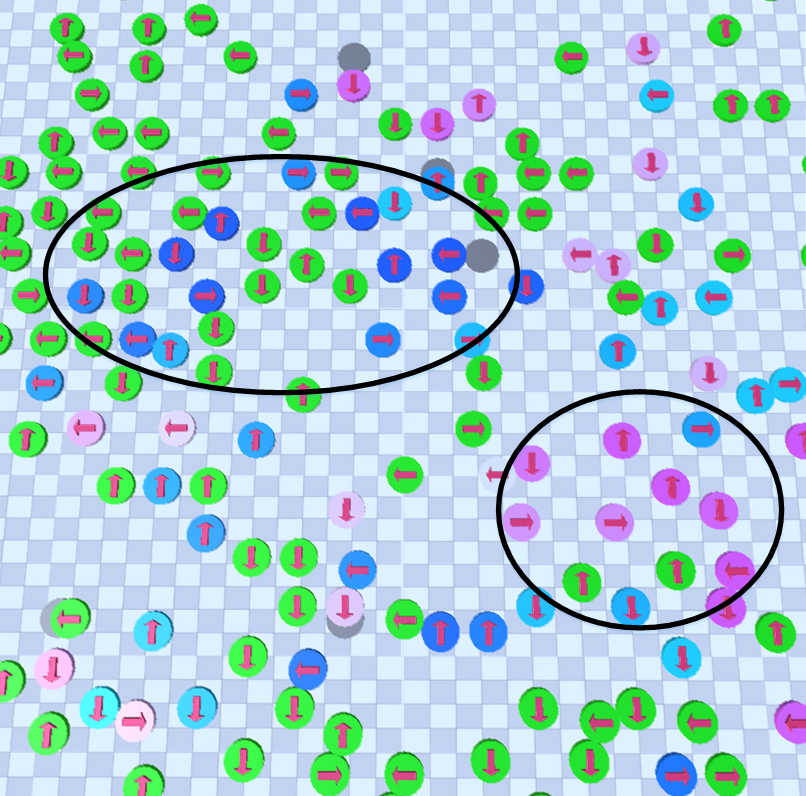} &  \includegraphics[width=3.5cm,height=3.5cm]{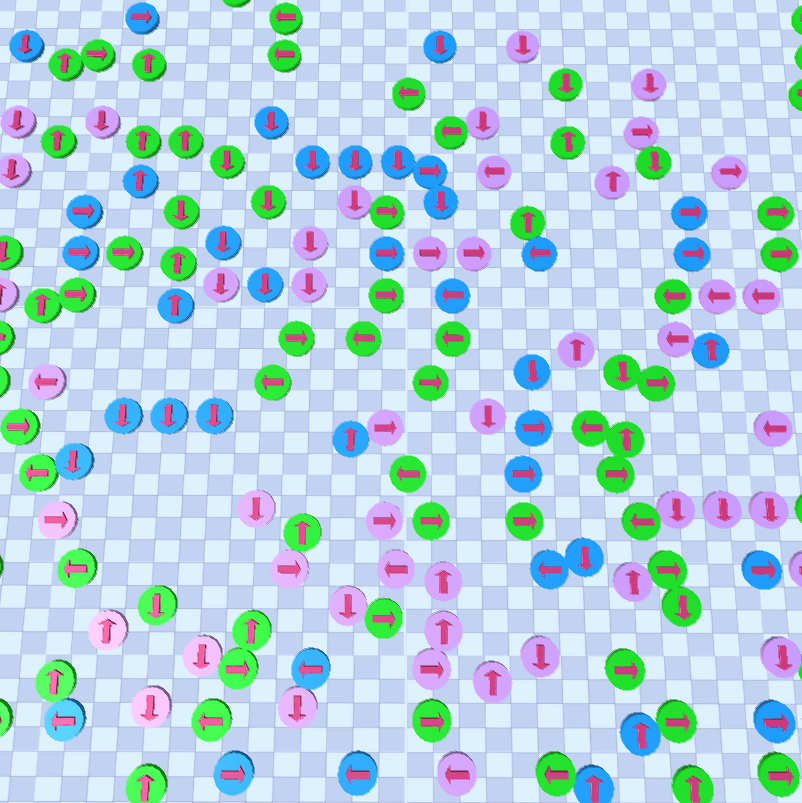} \\
 (a)   & (b)     \\
\end{tabular}
\caption{
Simulation results (a) considering emotional contagion  and (b) not considering emotional contagion. In (a) the deterrent forces of the agents are different and the same types of agents are more likely to gather together. In (b) the deterrent forces of all the
agents are the same and the same type of agents are dispersed.
}
\label{fig:23}
\end{figure}

\subsection{The impact of emotional contagion on the result of crowd violence}
\label{The impact of emotional contagion on the final result of group violence}

The impact of emotional contagion on the results of crowd violence is presented in Figure \ref{fig:22}.
The initial numbers of civilians, activists, and cops are 80, 50, and 40, respectively. In Figure \ref{fig:22}a, the initial emotion
values of the cops are higher than those of the activists. At the 53rd time step, the number of activists is zero. All the
activists have been wiped out by the cops. At first the number of civilians increases because the total deterrent force
of the cops is much higher than that of the activists. Some activists change their role (from activist to civilian).
In Figure \ref{fig:22}b, all the agents without emotion have the same deterrent force. At the 41st time step, the number
of cops is zero and the activists win.


We can learn from Figure \ref{fig:22} that the emotion module of our model can describe the differences observed between the agents. In our model, the deterrent forces of all the agents are different according to their emotions.
The greater the absolute value of an agent's emotion, the higher the deterrent force of that agent.
Although the number of agents is small, it is still possible to overcome a larger group of opponents by improving the
emotions and deterrent forces of them.
Therefore, our model can simulate situations in which agents overcome their more numerous opponents.


Figure \ref{fig:23} shows simulation results with and without considering emotion. In Figure \ref{fig:23}a the deterrent forces of the agents
are different. The stronger the color intensity of a circle, the higher the deterrent force of the agent. Because of emotional
contagion, the same types of agents are more likely to gather together, which is similar to what happens in real-world scenarios \cite{031}. In Figure \ref{fig:23}b
the deterrent forces of all the agents are the same and the same type of agents are more dispersed.

\begin{figure}[htbp]
\centering
\begin{tabular}{cc}
 \includegraphics[width=3.2cm]{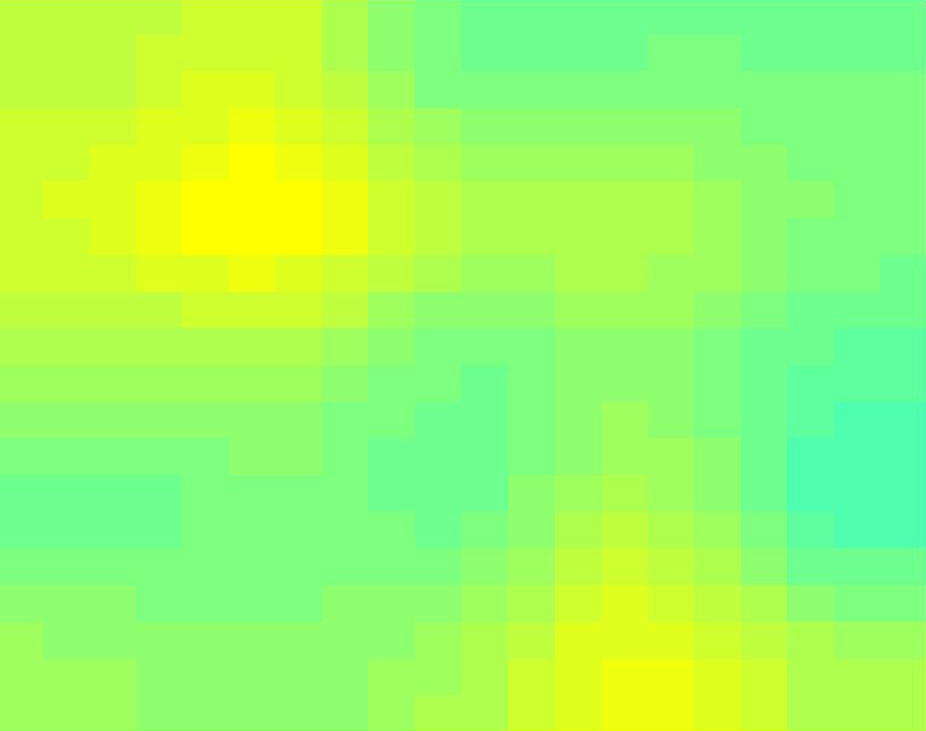} &  \includegraphics[width=3.2cm]{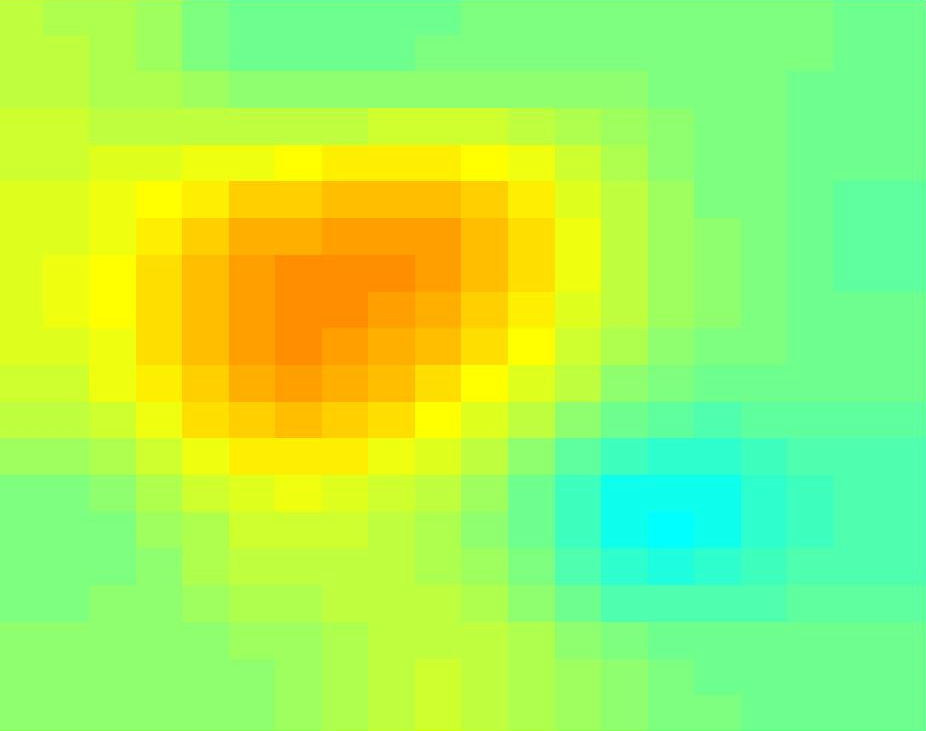} \\
 (a)   & (b)  \\
  \includegraphics[width=3.2cm]{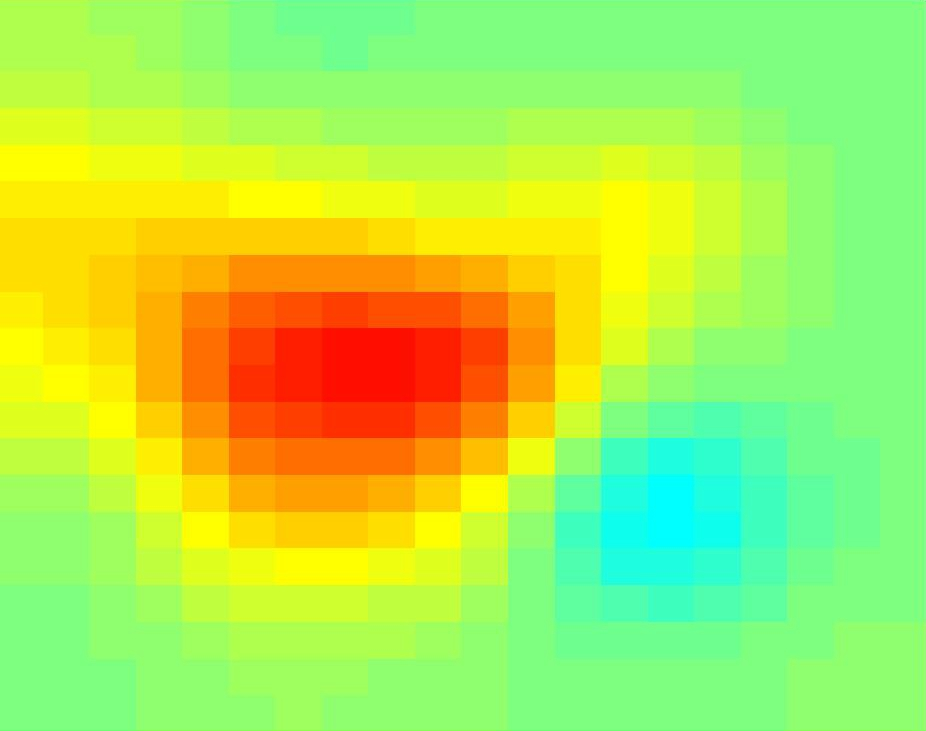} &  \includegraphics[width=3.2cm]{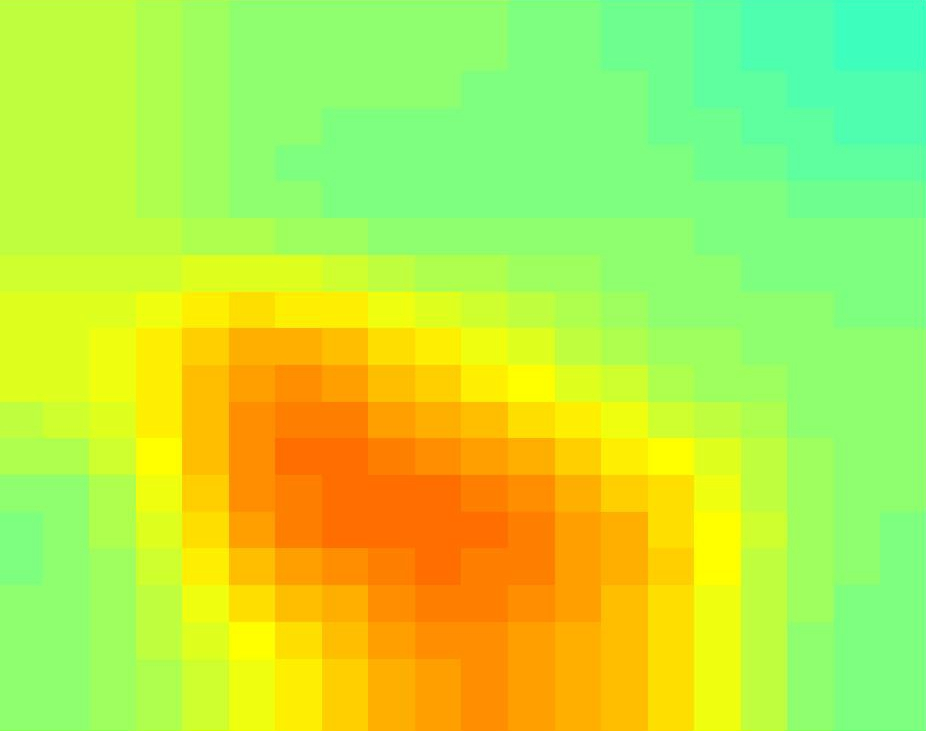} \\
 (c)  & (d)  \\
\end{tabular}
\caption{
The heat maps of antagonistic emotion: (a) heat map at the 5th time step, (b) heat map at the 17th time step, (c) heat map at the 27th time step, and (d) heat map at the 42th time step. The red area represents the emotions of the cops. The blue area represents the emotions of the activists. The stronger the color intensity, the larger the value of the emotion.
}
\label{fig:emo}
\end{figure}

Figure \ref{fig:emo} shows the heat maps of antagonistic emotion. Different colors represent different types of agents' emotions. The stronger the color intensity, the larger the value of the emotion. At first, the emotions of the cops and the activists are very weak. Later, as a result of the confrontation between cops and activists, both types of emotions increase. Since the initial emotions of the cops are higher than those of activists, the overall emotional scope of the cops becomes wider and wider and that of the activists becomes smaller and smaller. Finally, all the activists are wiped out by the cops and there is no blue area on the map.

\subsection{The influence of deterrent force on strategy selection}
\label{The influence of deterrent force on strategy}

\begin{figure}[t] \begin{centering}
  \centering
  \includegraphics[width=9cm]{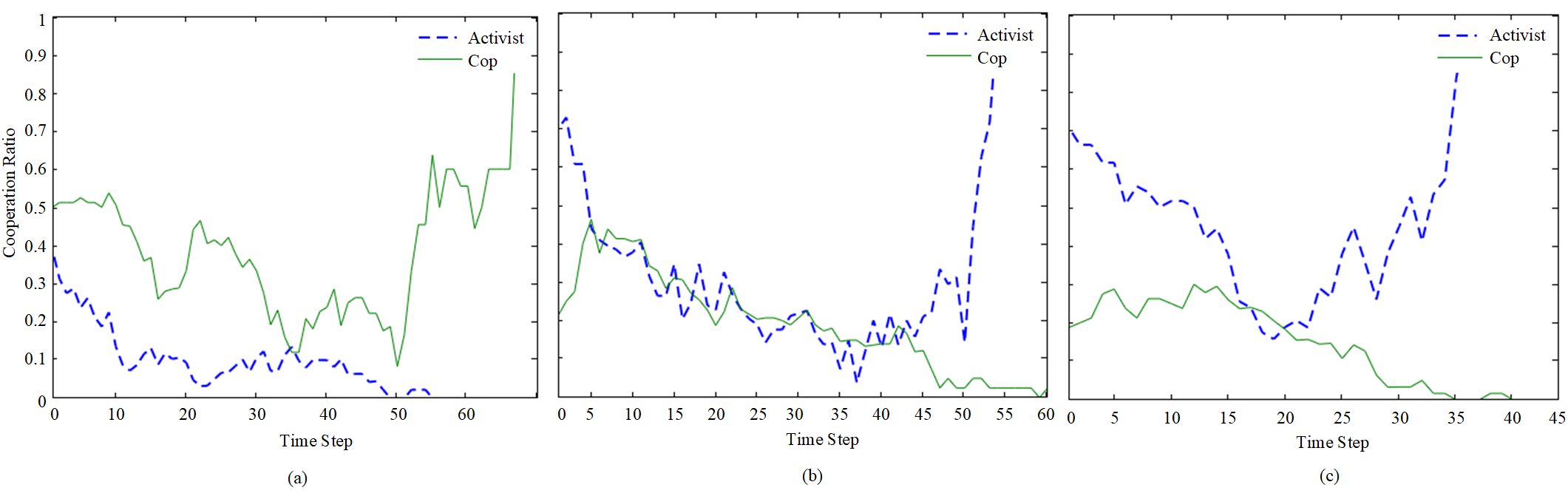}
  \centering
  \caption{ Cooperation ratios of cops and activists at different time steps:
(a) when the total deterrent force of the activists is higher than that of the cops;
(b) when the total deterrent force of the activists is equal to that of the cops;
(c) when the total deterrent force of the activists is less than that of the cops.}
  \label{fig:252627}
  \end{centering}
\end{figure}

We present the relationship between deterrent force and strategy selection in this subsection. We analyze the strategy (cooperation or defection)
adopted by each agent with respect to their different deterrent forces.


Figure \ref{fig:252627} shows the overview of cooperation ratios (the ratio of
the number of agents adopting a cooperation strategy to the total number of this type of agents) in relation to different deterrent forces.
When the total deterrent force of the activists is higher than that of the cops (Figure \ref{fig:252627}a), most
of the cops adopt a cooperation strategy to avoid causalities and conserve their fighting forces.
When the total deterrent force of the activists is equal to that of the cops, the cooperation ratios of
cops and activists are almost the same from the 5th time step to the 45th time step. After the 45th
time step, the cops defeat the activists. More and more cops adopt defection strategy, the cooperation
ratio of the cops decreases, and the cooperation ratio
of the activists increases. When the total deterrent force of the activists is less than that of the cops, the
cooperation ratio of the activists is higher than that of the cops.

In summary, the agent with a higher deterrent force is more likely to adopt a defection strategy and the agent with a lower deterrent
force is more likely to adopt a cooperation strategy.

\subsection{Comparisons}
\label{Comparisons}

\begin{figure}[htb]
\centering
\begin{tabular}{cc}
 \includegraphics[width=3.2cm]{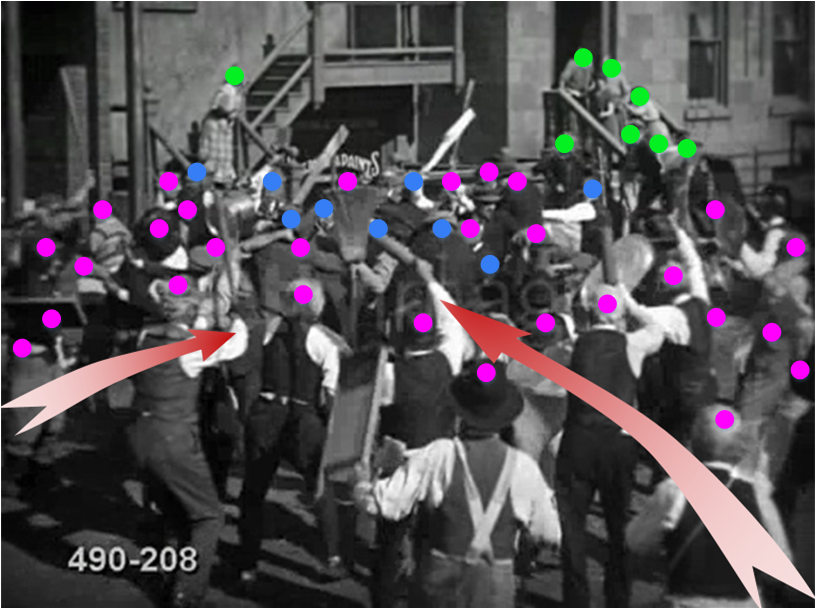} &  \includegraphics[width=3.2cm]{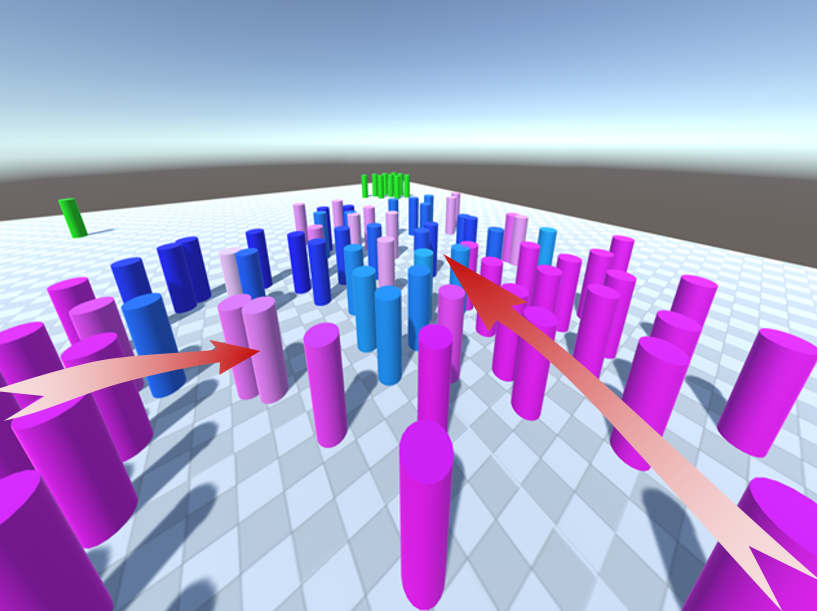} \\
 (a)   & (b)  \\
  \includegraphics[width=3.2cm]{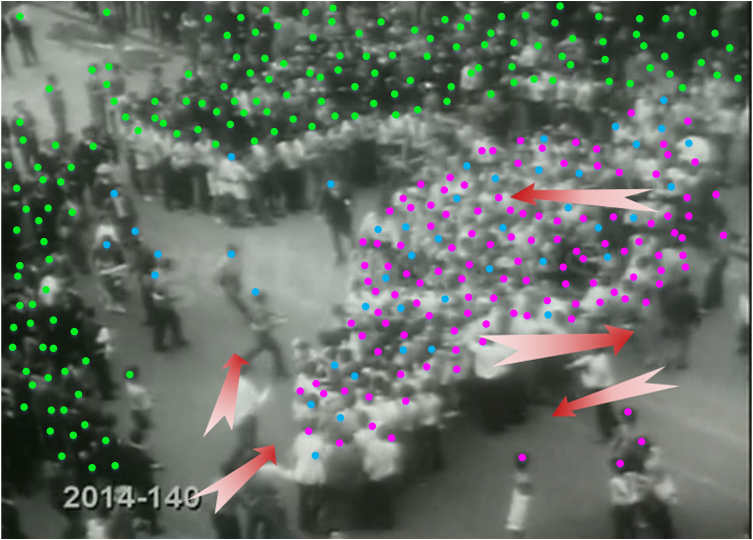} &  \includegraphics[width=3.2cm]{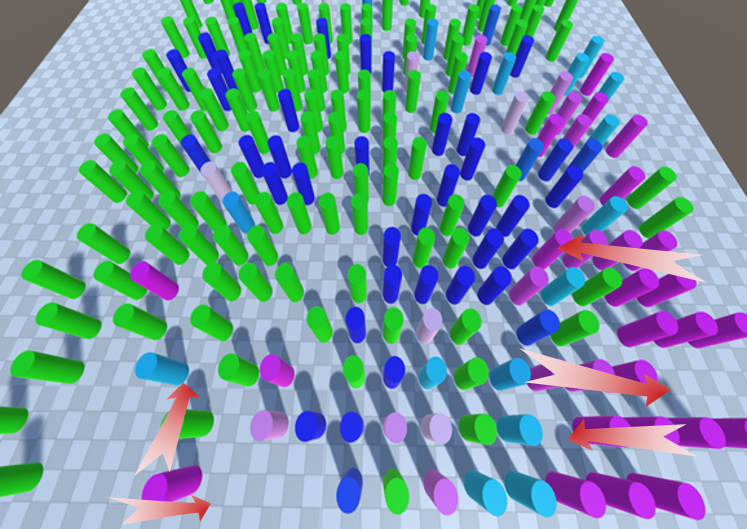} \\
 (c)  & (d)  \\
   \includegraphics[width=3.2cm]{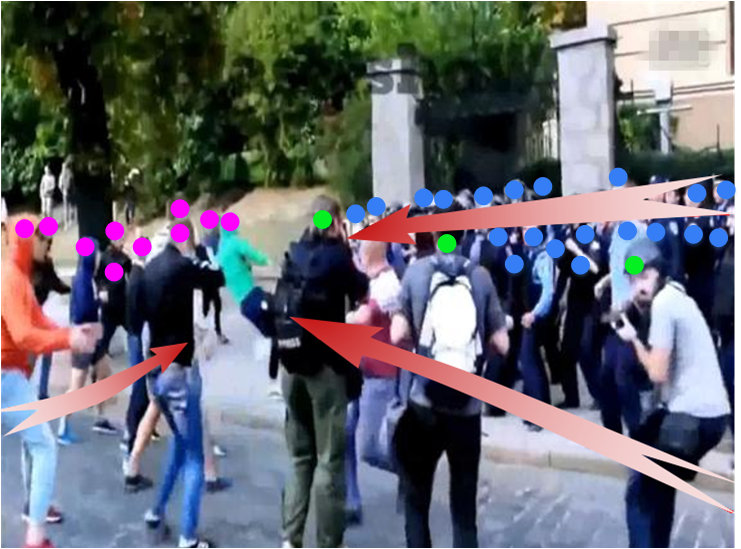} &  \includegraphics[width=3.2cm]{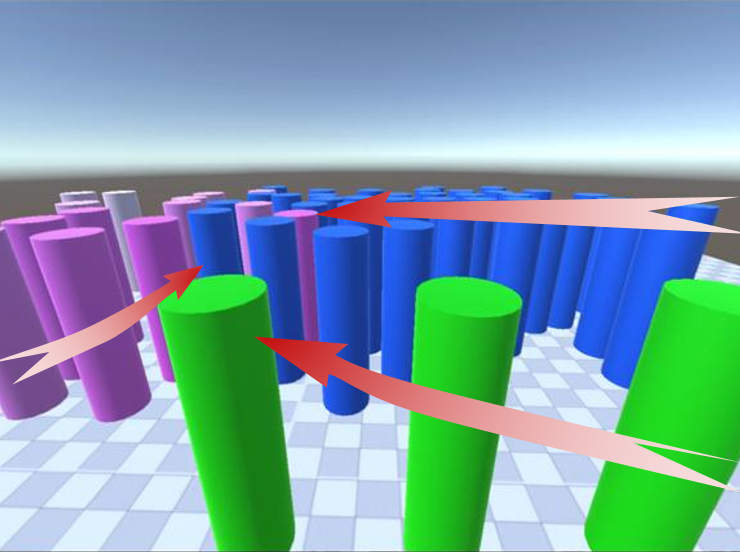} \\
 (e)  & (f)  \\
   \includegraphics[width=3.2cm]{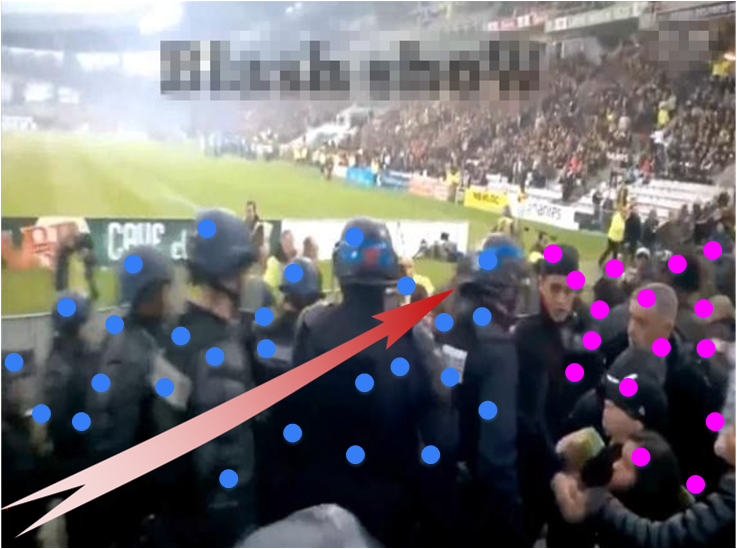} &  \includegraphics[width=3.2cm]{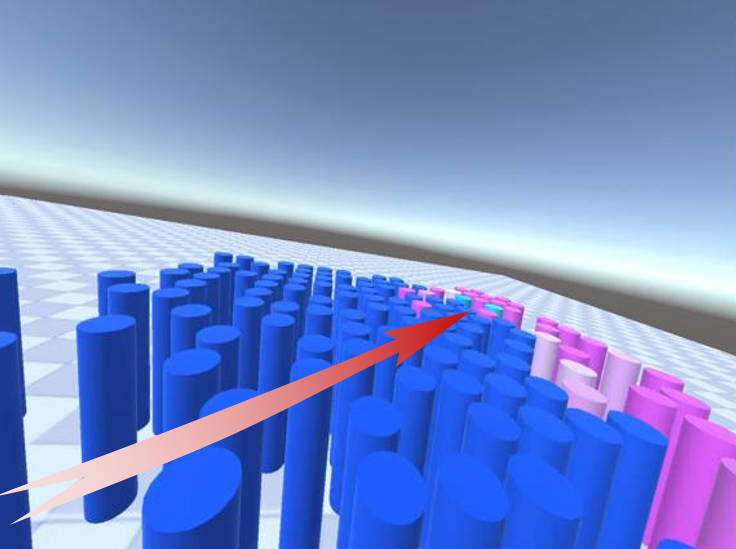} \\
 (g)  & (h)  \\
   \includegraphics[width=3.2cm]{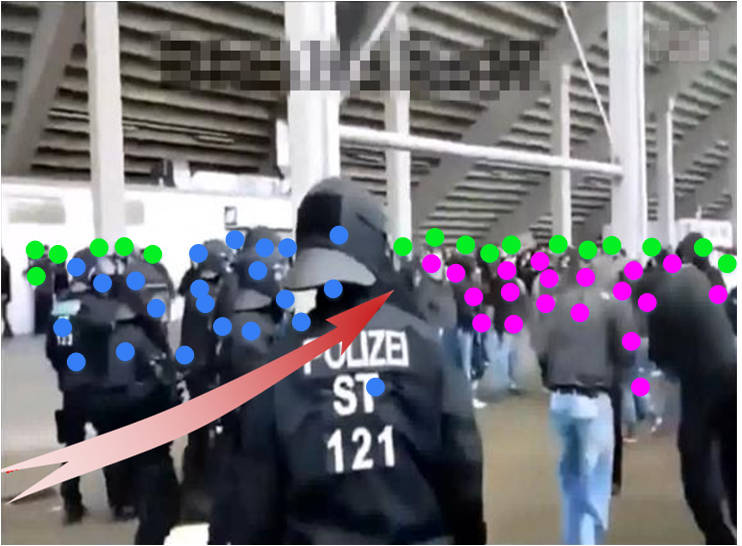} &  \includegraphics[width=3.2cm]{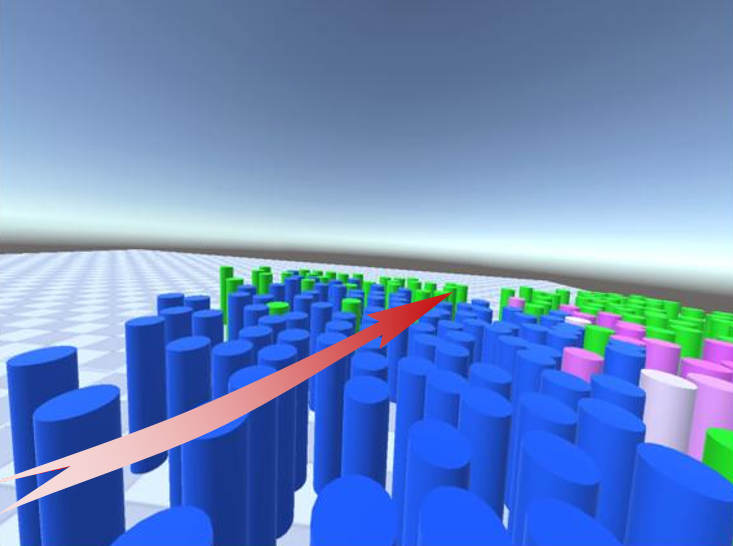} \\
 (i)  & (j)  \\
\end{tabular}
\caption{
Comparisons between real scenes and our simulation results. (a), (c), (e), (g), and (i) are five real antagonistic scenes. (b), (d), (f), (h), and (j) are our corresponding simulation results generated by our ACSEE model. The red arrows are the dominant paths of crowd movements.
}
\label{fig:28}
\end{figure}

To validate our approach, we compare our simulation results with those generated by the CVM \cite{027} and ABEC \cite{120} models and real-world videos.
Main goal of our model is to predict trends of crowd movements in these situations without explicitly modeling the trajectory of a particular individual. Simulation results obtained by our model are closest to the real-world scenes in the overall trends of crowd movements.

The real scenes shown in Figures \ref{fig:28}a and \ref{fig:28}c are chosen from the public web dataset \cite{071}. The real scenes of Figures \ref{fig:28}e, \ref{fig:28}g, and \ref{fig:28}i are chosen from real antagonistic incidents on YouTube.
For the real-world scenarios in Figure \ref{fig:28}, we use different colors to distinguish different roles of individuals. The green, purple, and blue circles and cylinders represent civilians, activists, and cops, respectively. The red arrows are the dominant paths of crowd movements.
Cops and activists in this case represent two opposing groups. Civilians in this case represent onlookers and neutrals.
There are no deaths of activists or cops in these scenes. The scenes can be simulated by increasing the values of the thresholds of $T_{warn}$ and $T_{warn\_time}$ for our model. The parameter values used in the simulation runs are listed in Table \ref{pv}.
More details can be seen in the supplementary video.

We use the dominant path and entropy metric to quantitively evaluate our simulation results. Dominant path is defined based on collectiveness of crowd movements. Collectiveness, which indicates the degree to which individuals acting as a unit, is a fundamental and universal measurement for various crowd systems, including crowds in antagonistic scenes \cite{097}. Individuals locally coordinate their movements and behaviors with their neighbors, and then the crowd is self-organized into collective movements without external control.
The method proposed in \cite{033} is used to calculate the collectiveness of our simulation results. When the collectiveness of individual movements in a certain area is significantly higher than that of surrounding area, a small group of agents with similar movements is formed. The center of the group is determined according to the average of the positions of all the agents in this group. The trajectory of the group center forms its dominant path and can be treated as the movement trend of the whole group. Using this method, we can get the dominant paths of the real-world videos.
We use entropy metric \cite{096}, angular error (AE) \cite{122}, and inter-group distance metric (IDM) \cite{121} to evaluate our crowd simulation results. They are used to evaluate the errors of trajectories, movement directions, and distances between each group, respectively. These three evaluation methods complement each other, which are used to evaluate our results more comprehensively.

An entropy metric \cite{096} is adopted to evaluate the error between the dominant paths of simulation results and that of the real-world videos. A lower entropy value implies a higher similarity between the simulation results and the real-world scenarios. Table \ref{em5s} shows the entropy metric of the simulation results achieved by the CVM, ABEC, and ACSEE (ours) models on different scenarios in Figure \ref{fig:28}. The simulation results obtained by our model conform to the real-world videos best.


The angular error (AE) \cite{122} between the movement direction of the simulation result $(V_x, V_y)$ and that of the ground-truth $(V_{xgt}, V_{ygt})$ is also used to evaluate our crowd simulation method. This AE is defined in Equation \ref{mycos}. The inter-group distance metric (IDM) \cite{121} compares the difference in the average distances between each pair of clustered agents. Tables \ref{myae} and \ref{myide} show the AE and IDM of our simulation results on different scenarios in Figure \ref{fig:28}.

Tables \ref{em5s}, \ref{myae}, and \ref{myide} show that our method consistently outperforms the CVM and ABEC models. Compared with the CVM model, our ACSEE model considers the agents' antagonistic emotions and accurately describe the differences between agents' deterrent forces. Compared with the ABEC model only considering emotional contagion in antagonistic scenarios, our method considers not only antagonistic emotion, but also the relationship between antagonistic emotion and evolutionary game theory. The agent is able to choose a more reasonable strategy according to the situation.

\begin{equation}\label{mycos}\scriptsize
AE=cos^{-1} ((V_x \cdot V_{xgt}+V_y \cdot V_{ygt})/{\sqrt{V_x^2+V_y^2} \sqrt{V_{xgt}^2+V_{ygt}^2 }})
\end{equation} 

\begin{table}[htbp]
\setlength{\belowcaptionskip}{10pt}
\centering
\caption{Entropy metric for our simulation algorithms on different scenarios from Figure \ref{fig:28}. A lower entropy value implies higher similarity between the simulation results and the real-world scenarios. Simulations with an entropy score less than $1.000$ are considered very visually similar to the source data and those with a score greater than $6.000$ are visually very different \cite{096}. Scene No. 2 (Figure \ref{fig:28}c) is too large and chaotic and the collectiveness of crowd movement is not so obvious. Therefore, the entropy value of Scene No. 2 is larger than $1.000$, but this value is far less than $6.000$. }
\begin{tabular}{c|c|c|c|l|c}
\hline
Scene No.      & 1     & 2     & 3     & \multicolumn{1}{c|}{4} & 5     \\ \hline\hline
ACSEE & \textbf{0.193} & \textbf{1.310} & \textbf{0.255} & \textbf{0.104}                  & \textbf{0.117} \\ \hline
ABEC & 0.232 &1.320 & 0.268 & 0.139 & 0.180 \\ \hline
CVM & 0.259 & 1.369 & 0.270 & 0.151 & 0.187\\  \hline
\end{tabular}
\label{em5s}
\end{table}

\begin{table}[htbp]\tiny
\setlength{\belowcaptionskip}{10pt}
\centering
\caption{AE for the simulation algorithms on different scenes from Figure \ref{fig:28}. A lower value for AE implies higher similarity with respect to the real-world crowd videos. We report mean and variance of AE at different time steps.  }
\begin{tabular}{c|c|c|c|l|c}
\hline
Scene No.      & 1     & 2     & 3     & \multicolumn{1}{c|}{4} & 5     \\ \hline\hline
ACSEE & \textbf{0.132/0.150} & \textbf{0.285/0.329} & \textbf{0.210/0.099} & \textbf{0.090/0.003}                  & \textbf{0.036/0.003} \\ \hline
ABEC & 0.214/0.155 & 0.463/3.027 & 0.999/12.822 & 0.500/0.002 & 0.558/0.006\\ \hline
CVM & 1.329/20.709 & 1.579/1.500 & 1.464/14.452 & 0.948/0.007 & 0.856/0.062\\  \hline
\end{tabular}
\label{myae}
\end{table}

\begin{table}[htbp]
\setlength{\belowcaptionskip}{10pt}
\centering
\caption{ IDM (pixel) for our simulation algorithms on different scenes from Figure \ref{fig:28}. Lower numbers are better.  }
\begin{tabular}{c|c|c|c|l|c}
\hline
Scene No.      & 1     & 2     & 3     & \multicolumn{1}{c|}{4} & 5     \\ \hline\hline
ACSEE & \textbf{1} & \textbf{30} & \textbf{2} & \textbf{6}  & \textbf{12} \\ \hline
ABEC & 31 & 32 & 21 & 18 & 35\\ \hline
CVM & 40 & 42 & 32 & 49 & 63\\  \hline
\end{tabular}
\label{myide}
\end{table}




The simulation result generated by our model is compared with those of the CVM \cite{027} and ABEC \cite{120} models in Figure \ref{fig:29}.
More details can be seen in the supplementary video.
In contrast to the CVM and ABEC models, our method can better quantify the differences of the deterrent forces of all the agents. Besides, we also integrate antagonistic emotional contagion into evolutionary game theory to estimate situations of antagonistic scenes more accurately, which helps agents make reasonable strategies in the games. Therefore our simulation result is the most similar to the real-world scenario.

\subsection{User Studies}
\label{User Studies}


In this section we describe user studies conducted to demonstrate the perceptual benefits of our ACSEE model compared to other models in simulating antagonistic crowd behaviors.


\textbf{Experiment Goals $\&$ Expectations}: Our main goal is to measure how close the crowd movement tendencies generated using different models are to those observed in real-world videos. We hypothesize that in both studies, agents simulated with our ACSEE model will exhibit overall more plausible antagonistic crowd movements than other models. Therefore, participants will strongly prefer our model to the other models.


\textbf{Experimental Design}: Two user studies were conducted based on a paired-comparison design. In each study, participants were shown pre-recorded videos in a side-by-side comparison of simulation results generated by different models and a real-world video. In particular, we asked the users to compare the crowd movement tendencies generated by different crowd simulation models with those observed in real-world videos. The studies had no time constraints and the participants were free to take breaks between the benchmarks. We encouraged the users to watch these videos as many times as they wanted and finally give a stable score.


\textbf{Comparison Methods}: The first study compares our ACSEE model considering emotion with our model that does not consider emotion. The second study compares our model with the CVM \cite{027} and ABEC \cite{120} models.

\begin{figure}[t]
\centering
\begin{tabular}{cc}
 \includegraphics[width=3.2cm,height=2.26468cm]{scene3real.jpg} &  \includegraphics[width=3.2cm,height=2.26468cm]{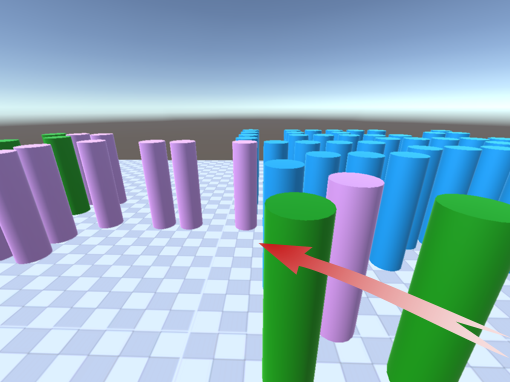} \\
 (a) Real scenario  & (b) CVM result \\
  \includegraphics[width=3.2cm,height=2.26468cm]{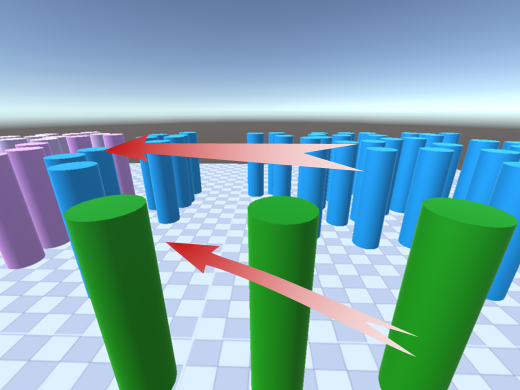} &  \includegraphics[width=3.2cm,height=2.26468cm]{scene3sim.jpg} \\
 (c) ABEC result  & (d) Our result \\
\end{tabular}
\caption{
Comparisons between the real scenario and simulation results of different crowd simulation models.
The simulation result obtained by our model conform to the real-world video best.
}
\label{fig:29}
\end{figure}

\begin{figure}[t] \begin{centering}
  \centering
  \includegraphics[width=7cm]{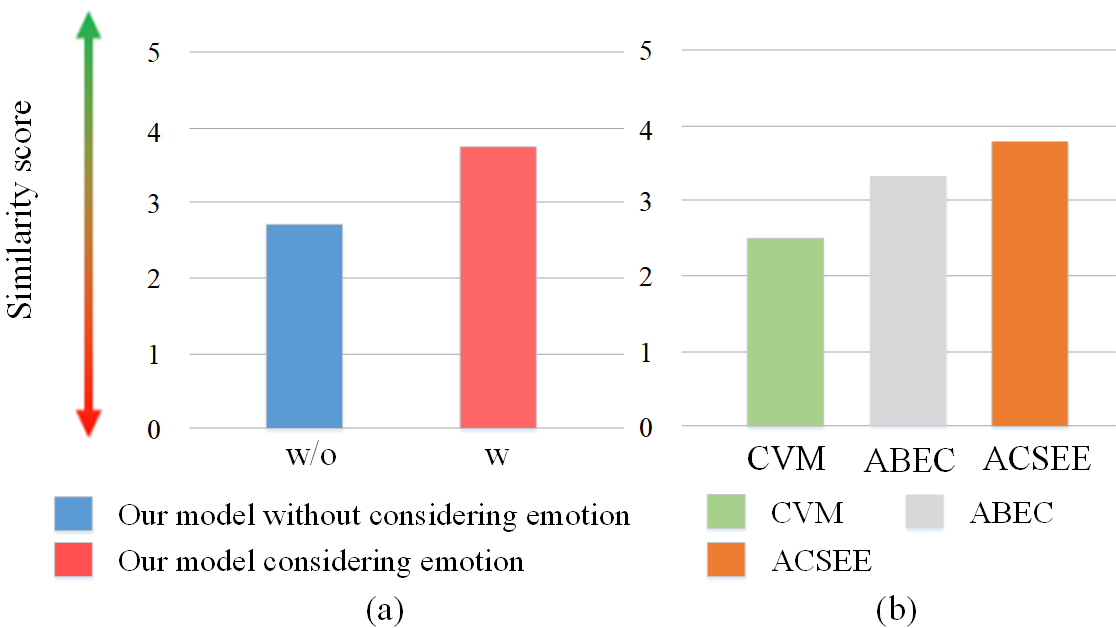}
  \centering
  \caption{
  User evaluation of simulation scenes. (a) Comparison of simulation results considering emotion and without considering emotion. (b) Comparison of simulation results generated by the CVM, ABEC, and ACSEE models. Participants were asked to rate each simulation result on a scale of 1 - 5 in terms of the similarity of movement tendencies between the real-world videos and the simulation result videos. We can see from the results that our simulation results are most similar to real-world scenarios.
}
  \label{fig:222}
  \end{centering}
\end{figure}



\textbf{Environments}: We use outdoor scenarios without obstacles. In these scenarios, the green, purple, blue, and grey circles are civilians, activists, cops, and dead agents, respectively.


\textbf{Metrics}: Participants were shown two pre-recorded videos in a side-by-side comparison of the simulation result and a real-world video. We asked the users to first watch the real-world video and then rate each simulation result on a scale of 1 - 5 in terms of the similarity of movement tendencies between the real-world video and the simulation result video. A score of 1 indicates most dissimilar and a score of 5 indicates most similar movement tendencies.


\textbf{Results}: There are 39 participants (20 female) with a mean age of $25.12 \pm 3.26$ years in these studies. We measure the mean and the variance of their scores and then compute the p-values using a two-tailed t-test. The means of their scores for our model without emotion and with emotion are $2.87 \pm 0.67$ and $3.73 \pm 0.93$, respectively. The means of their scores for the CVM, ABEC, and ACSEE models are $2.56 \pm 1.44$, $3.30 \pm 0.50$, and $3.70 \pm 0.90$, respectively. The p-value for Figure \ref{fig:222}a comparison is $2.97e^{-26}$. The p-values for the comparison of the CVM and ACSEE models and that of the ABEC and ACSEE models in Figure \ref{fig:222}b are $2.97e^{-26}$ and $1.05e^{-17}$, respectively. We observe that the antagonistic crowd behavior simulations generated by our ACSEE model score much higher than the other models at a statistically significant rate (p-value$<$0.05). The result indicates that the addition of antagonistic emotion and the integration of antagonistic emotion and evolutionary game theory improve the perceptual similarity of our simulations to the crowd movement tendencies in real-world scenes. Our model gets higher scores than the CVM and ABEC models, as detailed in Figure \ref{fig:222}b. Participants indicate their preference for our ACSEE model.

\section{Conclusion and Limitations}
\label{conclusion}

We present a new model for antagonistic crowd behavior simulation integrated with emotional contagion and evolutionary game theory. Our approach builds on well-known psychological theories to present a comprehensive and antagonistic emotional contagion model.
Based on the emotional calculation method, we propose the deterrent force to determine the situation of cops
and activists. According to the situation, an enhanced evolutionary
game theoretic approach incorporated with antagonistic emotional contagion is determined.
Finally, we present a behavior control decision method based on the antagonistic emotional
contagion and evolutionary game theoretic approaches.

Our proposed model is verified by simulations. We investigate the impact of
different factors (number of agents, emotion, strategy, etc.) on the outcome of crowd violence.
Our model is compared with real-world videos and previous approaches. Results show that our proposed model can reliably
generate realistic antagonistic crowd behaviors.

However, our model still has several limitations.
Although our simulation results are closer to the real-world scenes in the overall trend of crowd movement, the antagonistic emotions in a crowd violent scene cannot be obtained directly or inferred accurately.
One of the main reasons is that the quality of most of the real videos is poor, since they are often captured by moving phones. At present, there is no effective methods to identify and quantify the emotion values of all the individual in such videos with poor quality.
Thus, the initial state of our model is set empirically according to real-world videos, which is time-consuming and not very accurate.
In the future, we plan
to use the latest wearable equipment to collect these data and provide a new method that can quickly and accurately obtain the initial state.
Moreover, the strategies and benefits calculated by our antagonistic evolutionary game theoretic approach are the ideal situations.
Game theory assumes that all individuals are rational. However, some people in real scenes are irrational and extreme, which doesn't fully satisfy the precondition of game theory.
 In practice, people do not necessarily adopt the optimal strategy because of the limitations of perception and other complex factors.
Our current calculation result is optimal, which is only one of the possible results. In fact, it is impossible for all simulation results to be consistent with the real results.
We will continue to improve our prediction results considering more actual situations of antagonistic crowds.
At present, the behavior control in our model is proposed based on the cellular automata \cite{009}. Other more complex behavioral control methods will be further considered.

\section{Acknowledgments}
This work was supported by National Natural Science Foundation of China under Grant Number 61672469, 61772474, 61822701, 61872324, Program for Science \& Technology Innovation Talents in Universities of Henan Province (20HASTIT021, 18HASTIT020)  and Youth Talent Promotion Project in Henan Province (2019HYTP022).

%
%
%
%
%


%


\ifCLASSOPTIONcaptionsoff
  \newpage
\fi

\bibliographystyle{IEEEtran}
\bibliography{bare_jrnl_compsoc}

\begin{IEEEbiography}[{\includegraphics[width=1in,height=1.25in,clip,keepaspectratio]{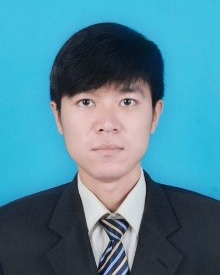}}]{Chaochao Li}
is a Ph.D candidate in the School of Information Engineering of Zhengzhou University, China, and his research interest is computer graphics and computer vision. He got his B.S. degree in computer science and technology from the School of Information Engineering of Zhengzhou University.
\end{IEEEbiography}
\begin{IEEEbiography}[{\includegraphics[width=1in,height=1.25in,clip,keepaspectratio]{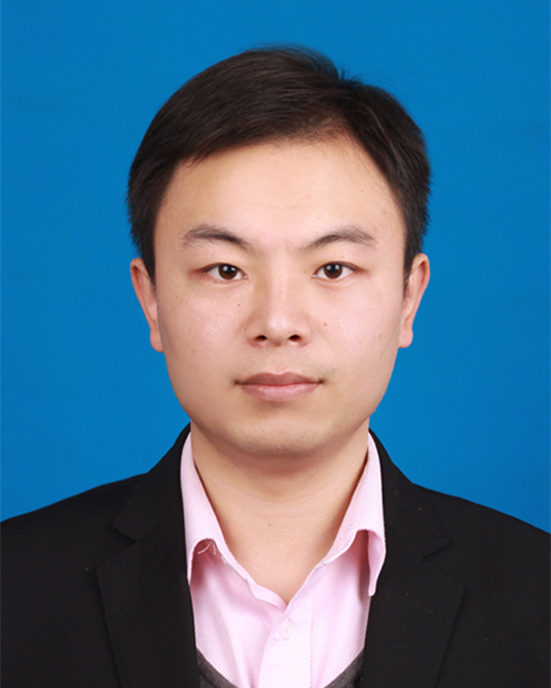}}]{Pei Lv}
 is an associate professor in School of Information Engineering, Zhengzhou University, China. His research interests include video analysis and crowd simulation. He received his Ph.D in 2013 from the State Key Lab of CAD\&CG, Zhejiang University, China. He has authored more than 20 journal and conference papers in these areas, including IEEE TIP, IEEE TCSVT, IEEE TAC、ACM MM, etc.
\end{IEEEbiography}

\begin{IEEEbiography}[{\includegraphics[width=1in,height=1.25in,clip,keepaspectratio]{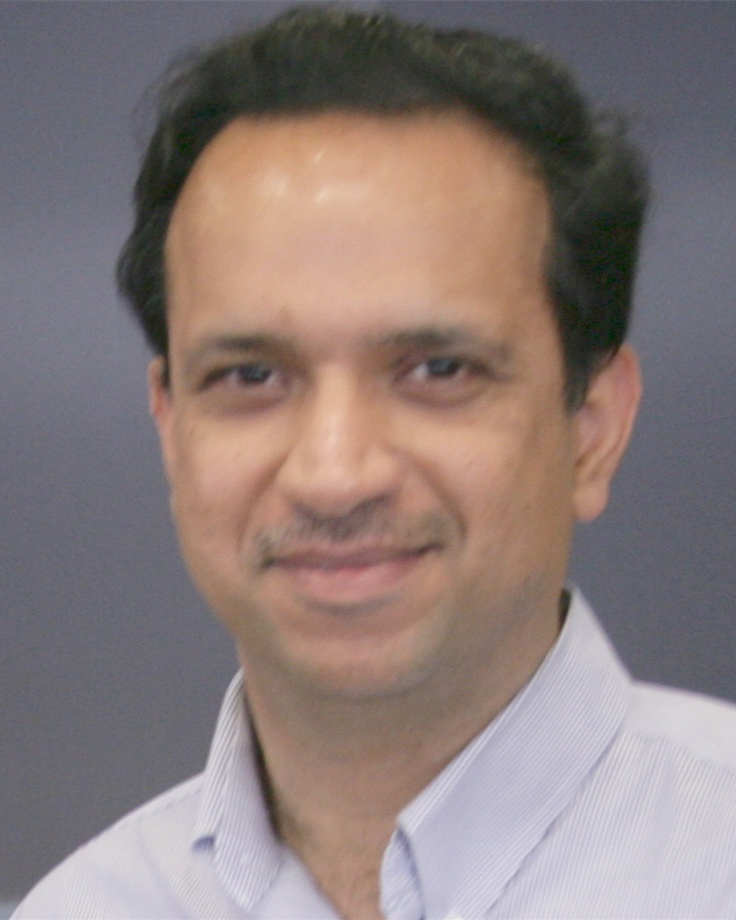}}]{Dinesh Manocha}
is the Paul Chrisman Iribe Chair in Computer Science \& Electrical and Computer Engineering at the University of Maryland College Park. He is also the Phi Delta Theta/Matthew Mason Distinguished Professor Emeritus of Computer Science at the University of North Carolina - Chapel Hill. He has won many awards, including Alfred P. Sloan Research Fellow, the NSF Career Award, the ONR Young Investigator Award, and the Hettleman Prize for scholarly achievement. His research interests include multi-agent simulation, virtual environments, physically-based modeling, and robotics. His group has developed a number of packages for multi-agent simulation, crowd simulation, and physics-based simulation that have been used by hundreds of thousands of users and licensed to more than 60 commercial vendors. He has published more than 510 papers and supervised more than 35 PhD dissertations. He is an inventor of 9 patents, several of which have been licensed to industry. His work has been covered by the New York Times, NPR, Boston Globe, Washington Post, ZDNet, as well as DARPA Legacy Press Release. He is a Fellow of AAAI, AAAS, ACM, and IEEE and also received the Distinguished Alumni Award from IIT Delhi. See \url{http://www.cs.umd.edu/~dm}
\end{IEEEbiography}
\begin{IEEEbiography}[{\includegraphics[width=1in,height=1.25in,clip,keepaspectratio]{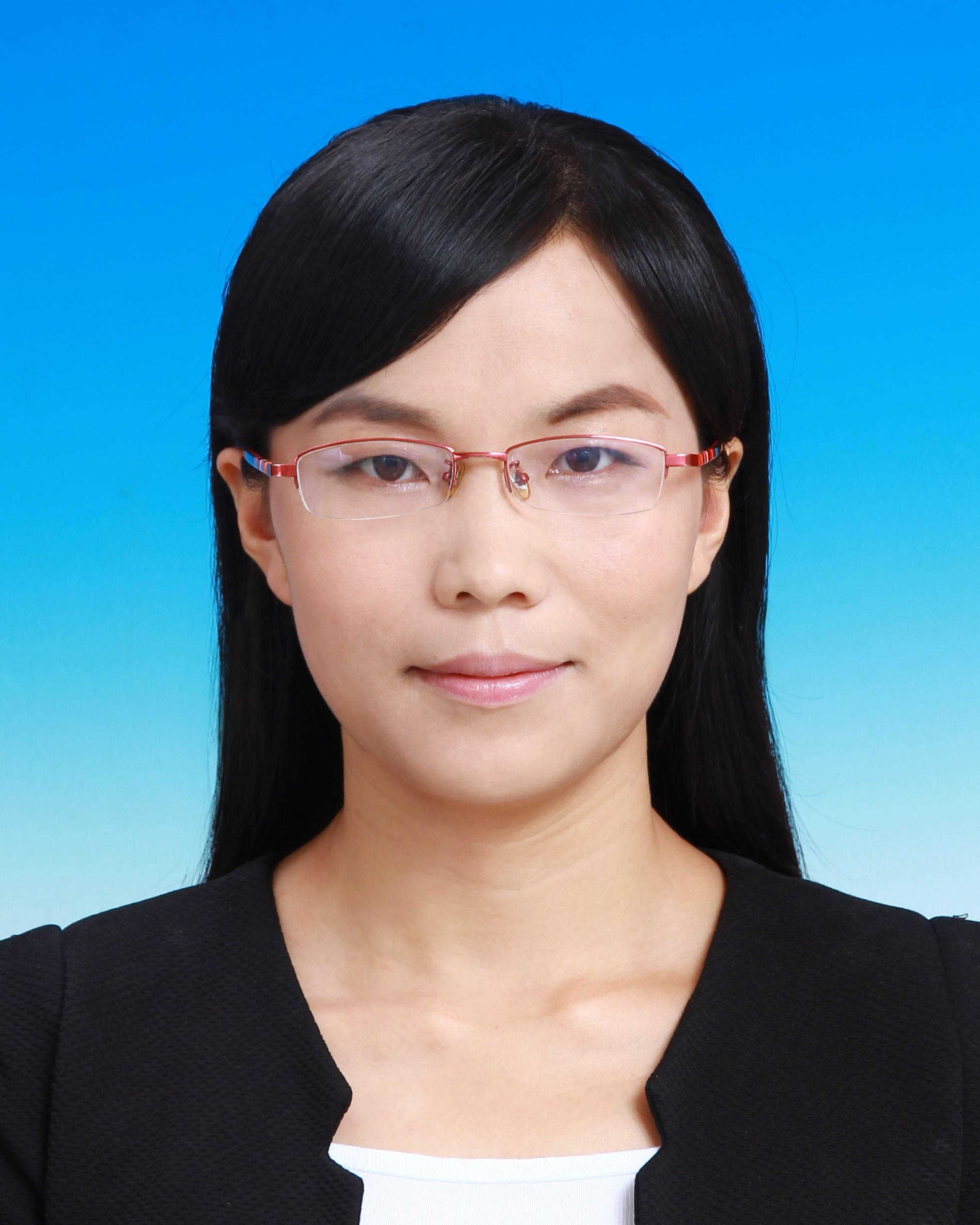}}]{Hua Wang}
was born in 1982. She received her PhD degree in Computer Science from the Institute of Computing Technology, Chinese Academy of Sciences, in 2015. She is currently an associate professor of Zhengzhou University of Light Industry, Zhengzhou, China. Her main research interests include traffic animation and environment modeling.
\end{IEEEbiography}
\begin{IEEEbiography}[{\includegraphics[width=1in,height=1.25in,clip]{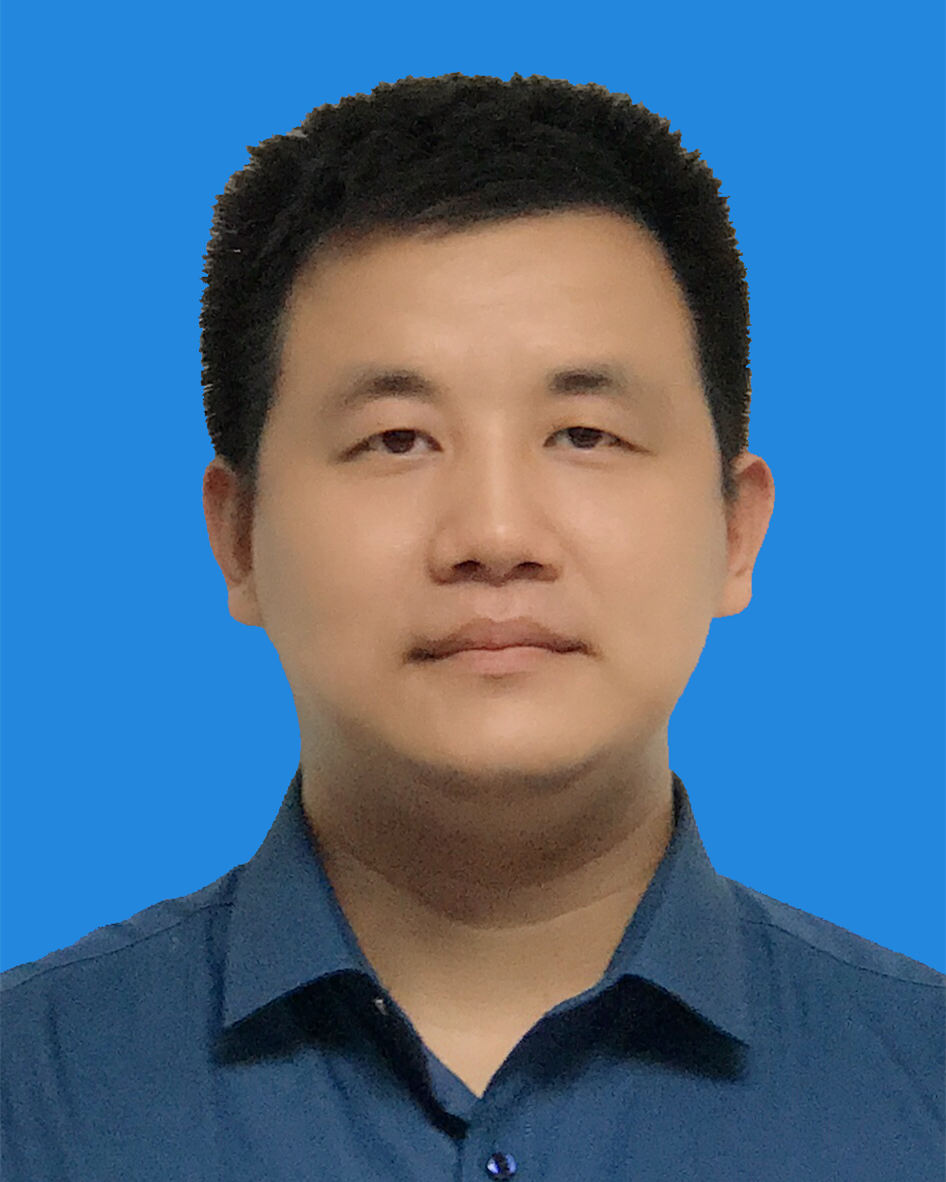}}]{Yafei Li}
is an assistant professor in the School of Information Engineering, Zhengzhou University, Zhengzhou, China. He received the PhD degree in computer science from Hong Kong Baptist University, in 2015. He held a visiting position in the Database Research Group with Hong Kong Baptist University. His research interests span mobile and spatial data management, location-based services, and urban computing. He has authored more than 20 journal and conference papers in these areas, including IEEE TKDE, IEEE TSC, ACM TWEB, ACM TIST, PVLDB, IEEE ICDE, WWW, etc.
\end{IEEEbiography}
\begin{IEEEbiography}[{\includegraphics[width=1in,height=1.25in,clip]{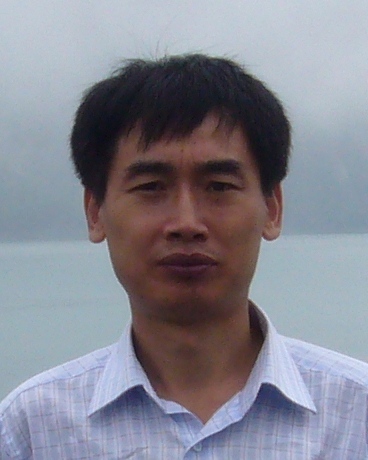}}]{Bing Zhou}
is currently a professor at the School of Information Engineering, Zhengzhou University, Henan, China. He received the B.S. and M.S. degrees from Xi'an Jiao Tong University in 1986 and 1989, respectively,and the Ph.D. degree in Beihang University in 2003, all in computer science. His research interests cover video processing and understanding, surveillance, computer vision, multimedia applications.
\end{IEEEbiography}
\begin{IEEEbiography}[{\includegraphics[width=1in,height=1.25in,clip,keepaspectratio]{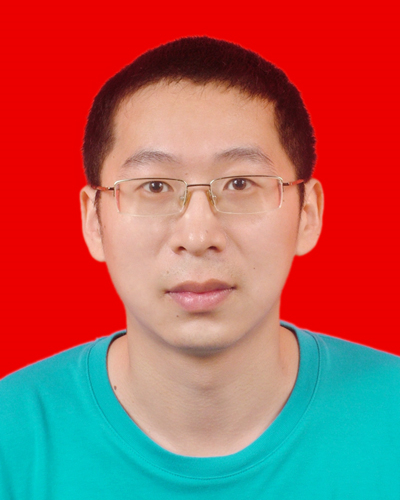}}]{Mingliang Xu}
is a full professor in the School of Information Engineering of Zhengzhou University, China, and currently is the director of CIISR (Center for Interdisciplinary Information Science Research) and the vice General Secretary of ACM SIGAI China. He received his Ph.D. degree in computer science and technology from the State Key Lab of CAD\&CG at Zhejiang University, Hangzhou, China. He previously worked at the department of information science of NSFC (National Natural Science Foundation of China), Mar.2015-Feb.2016. His current research interests include computer graphics, multimedia and artificial intelligence. He has authored more than 60 journal and conference papers in these areas, including ACM TOG, ACM TIST, IEEE TPAMI, IEEE TIP, IEEE TCYB, IEEE TCSVT, ACM SIGGRAPH (Asia), ACM MM, ICCV, etc.
\end{IEEEbiography}

\end{document}